\documentclass[preprint,journal]{vgtc}            % preprint (journal style)

\onlineid{0}

\vgtccategory{Research}

\title{Conversational Tactile Data Interfaces: Co-Designing Accessible Data Experiences with Blind Users Using Refreshable Tactile Displays and Conversational AI}

\author{%
  Samuel Reinders, 
  Munazza Zaib,
  Bongshin Lee, 
  Ingrid Zukerman,
  Matthew Butler,  \\
  Thien Autran,
  Sascha Cowley, 
  Francois Jacobs, 
  Lizhen Qu, and
  Kim Marriott
}

\authorfooter{
  \item
  	Sam Reinders, Munazza Zaib, Ingrid Zukerman, Matthew Butler, Lizhen Qu, and Kim Marriott are with Monash University.
  	E-mail: first.last@monash.edu
  \item
  	Bongshin Lee is with Yonsei University.
  	E-mail: b.lee@yonsei.ac.kr
  \item
    T. Autran, S. Cowley, and F. Jacobs contributed as co-designers.
}

\abstract{Combining refreshable tactile displays (RTDs) with conversational AI offers a promising approach to accessible data visualization for people who are blind or have low vision (BLV). However, 
it remains an open question how these modalities should be integrated to support accessible data experiences. We address this through a co-design process with three BLV co-designers. Building on our prior Wizard-of-Oz study, we created a conversational tactile data interface (CTDI) that combines an RTD with an LLM-powered conversational agent, refined through four workshops over eight months. In addition to the resulting system, \name, we contribute design knowledge and recommendations for CTDIs. Co-designers used touch as the primary sensemaking channel for spatial understanding of the data's shape, trends, and relationships, reserved the agent for what touch could not resolve (\eg\ calculation and analysis), and used the chart on the RTD to verify the agent's responses. Key findings include: a layered presentation that scaffolds chart exploration through progressive, interactive layers; a feedback grammar that distinguishes user- and agent-initiated tactile feedback; and a sequential interaction pattern---select, confirm, ask, verify---where each step grounds the last.}

\keywords{Accessible data visualization, refreshable tactile displays, conversational agents, interactive data exploration, co-design, people who are blind or have low vision.}

\teaser{
  \centering
  \includegraphics[width=\linewidth]{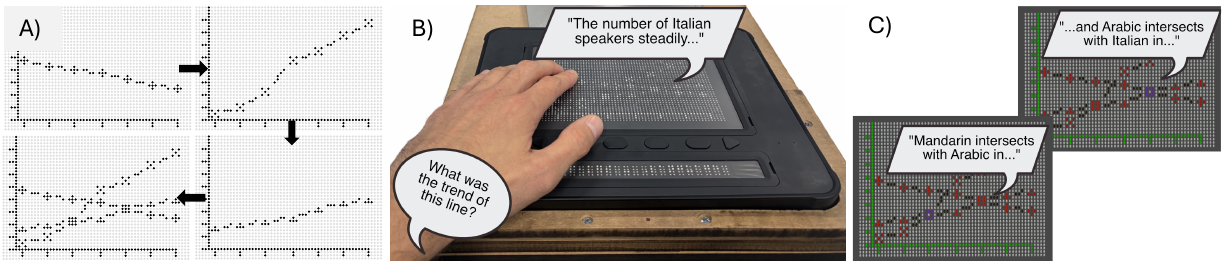}
  \caption{We co-designed \name, a conversational tactile data interface combining a refreshable tactile display with an LLM-powered conversational agent. A)~The layered presentation introduces chart components one at a time, each rendered in isolation on the RTD with a spoken description; B)~A user explores a line chart by touch, querying the agent with deictic references to their selected points; C)~Agent responses are segmented into navigable chunks synchronized with animated tactile highlights (shown in purple).}
  \label{fig:teaser}
}

\graphicspath{{figs/}{figures/}{pictures/}{images/}{./}} % where to search for the images

\usepackage{tabu}                      % only used for the table example
\usepackage{booktabs}                  % only used for the table example
\usepackage{lipsum}                    % used to generate placeholder text
\usepackage{mwe}                       % used to generate placeholder figures
\usepackage{ccicons}                   % package to be able to use icons from creative commons
\usepackage{comment}
\usepackage[percent]{overpic}
\usepackage{float}

\usepackage{mathptmx}                  % use matching math font
\usepackage{xcolor}
\usepackage{tikz}
\usepackage{enumitem}
\usetikzlibrary{positioning, calc, fit, backgrounds}
\definecolor{palegreen}{RGB}{200,230,200}
\definecolor{paleorange}{RGB}{255,229,204}
\definecolor{palepurple}{RGB}{255,204,255}
\definecolor{mintgreen}{RGB}{184,230,184}

\newcommand{\name}{\textit{Graphy}}
\newcommand{\etal}{{\em et al.}}
\newcommand{\eg}{e.g.,}

\newcommand{\bem}{\bfseries \itshape }

\newcommand{\fevol}[1]{\multicolumn{5}{c}{%
\begin{tikzpicture}[baseline=-0.5ex, x=11mm]
\useasboundingbox (-0.15,-0.15) rectangle (4.15,0.15);
#1
\end{tikzpicture}}}

\newcommand{\patA}{\fevol{\fill (0,0) circle (2pt); \draw[line width=1pt] (0,0) -- (1,0);}}
\newcommand{\patB}{\fevol{\fill (0,0) circle (2pt); \draw[line width=1pt] (0,0) -- (4,0);}}
\newcommand{\patC}{\fevol{\fill (0,0) circle (2pt); \draw[line width=1pt] (0,0) -- (4,0); \fill (1,0) circle (2pt);}}
\newcommand{\patE}{\fevol{\fill (1,0) circle (2pt); \draw[line width=1pt] (1,0) -- (4,0);}}
\newcommand{\patH}{\fevol{\fill (2,0) circle (2pt); \draw[line width=1pt] (2,0) -- (4,0);}}
\newcommand{\patI}{\fevol{\fill (2,0) circle (2pt); \draw[line width=1pt] (2,0) -- (4,0); \fill (3,0) circle (2pt);}}
\newcommand{\patJ}{\fevol{\fill (3,0) circle (2pt); \draw[line width=1pt] (3,0) -- (4,0);}}
\newcommand{\patK}{\fevol{\draw[line width=1pt, dashed] (1,0) -- (2,0); \fill[white] (1,0) circle (3.1pt); \draw[line
width=0.6pt] (1,0) circle (2pt); \fill (2,0) circle (2pt); \draw[line width=1pt] (2,0) -- (4,0);}}
  \newcommand{\patL}{\fevol{\draw[line width=1pt, dashed] (1,0) -- (2,0); \fill[white] (1,0) circle (3.1pt); \draw[line
  width=0.6pt] (1,0) circle (2pt); \fill (2,0) circle (2pt); \draw[line width=1pt] (2,0) -- (4,0); \fill (3,0) circle (2pt);}}

  \newcommand{\patM}{\fevol{\draw[line width=1pt, dashed] (2,0) -- (3,0); \fill[white] (2,0) circle (3.1pt); \draw[line
  width=0.6pt] (2,0) circle (2pt); \fill (3,0) circle (2pt); \draw[line width=1pt] (3,0) -- (4,0);}}

  \newcommand{\patO}{\fevol{\fill (0,0) circle (2pt); \draw[line width=1pt] (0,0) -- (4,0); \fill (2,0) circle (2pt);}}

\newcolumntype{Y}{>{\centering\arraybackslash}p{0.7cm}}

\newcommand{\bpstart}[1]{\smallskip\noindent{\textbf{#1.}}}
\newcommand{\bipstart}[1]{\smallskip\noindent{\textbf{#1.}}}
\newcommand{\myitem}{\vspace*{-1mm}\item}

\usepackage[normalem]{ulem}
\newif\ifmarkup
\markuptrue

\begin{document}

%%%%%%%%%%%%%%%%%%%%%%%%%%%%%%%%%%%%%%%%%%%%%%%%%%%%%%%%%%%%%%%%
%%%%%%%%%%%%%%%%%%%%%% START OF THE PAPER %%%%%%%%%%%%%%%%%%%%%%
%%%%%%%%%%%%%%%%%%%%%%%%%%%%%%%%%%%%%%%%%%%%%%%%%%%%%%%%%%%%%%%%

%% The ``\maketitle'' command must be the first command after the
%% ``\begin{document}'' command. It prepares and prints the title block.
%% the only exception to this rule is the \firstsection command
\firstsection{Introduction}

\maketitle

\label{sec:intro}
Data visualizations are increasingly accessed in all facets of everyday life: business, news, and education. For people who are blind or have low vision (BLV), accessing these data visualizations poses significant barriers. Existing approaches, like textual descriptions~\cite{jung2021communicating,Lundgard2022}, sonification~\cite{Kramer2010,holloway2022infosonics}, screen readers~\cite{Zong2022}, and static tactile graphics, each address aspects of accessible data visualization, but none provide the interactive, multimodal experience that lets BLV users independently explore, query, and verify data during sensemaking. 

Refreshable tactile displays (RTDs) render tactile graphics dynamically on pin-based displays, offering a more interactive alternative to static printed graphics, which require a new graphic to be produced for each change. However, RTDs in isolation have limitations: they are currently low resolution, which constrains labeling, and tactile graphics are difficult to comprehend without descriptions~\cite{BANA2010guidelines}. Combining an RTD with a conversational AI agent could address many barriers BLV people face when analyzing data~\cite{Chundury2026}: the agent provides verbal context and analytical support, while the rendered chart provides interactive spatial grounding and independent tactile access to the data. 

In our previous work, we conducted a Wizard-of-Oz (WOz) study revealing that BLV participants strongly preferred multimodal interaction combining touch with conversational queries~\cite{Reinders2024}. However, it remains an open question how these modalities should work together in a fully functional implementation. The wizard simulated an idealized system, seamlessly bridging touch and speech in real time, and abstracting away design challenges that a real system must solve, such as how to communicate feedback to users whose fingers must actively search for it, what role the agent should play relative to what touch already provides, and how to coordinate touch and speech to support chart orientation, exploration, sensemaking, and deictic interaction.

To address these questions, we co-designed \name\ with three BLV co-designers. \name\ is a \textbf{conversational tactile data interface (CTDI)}---a system combining an RTD with an LLM-powered conversational agent---that enables BLV users to explore data visualizations through touch and speech. Through four co-design workshops, the work presents an initial exploration of the design space for CTDIs. Our contributions are:

\begin{itemize}
    \myitem \textit{\name---the first conversational tactile data interface}, combining a multi-line RTD with an LLM-powered agent, created with three BLV co-designers. \name\ extends our prior WOz work into an instantiated system, revealing challenges WOz abstracted away and capabilities that emerged through co-design.
    
    \myitem \textit{Design knowledge for CTDIs}. Key findings include: a layered presentation that scaffolds chart exploration through interactive layers; a feedback grammar that distinguishes user- and agent-initiated tactile highlights; and a sequential interaction pattern---\textit{select, confirm, ask, verify}---grounding each step in the data. We distill these into design recommendations.   
\end{itemize}

\section{Related Work}
\label{sec:related}
\subsection{Accessible Data Visualization}
Accessible data visualization has attracted a growing research interest~\cite{lee2020reaching,kim2021accessible,marriott2021inclusive,dagstuhl2023inclusive}. Printed tactile graphics have long been the traditional approach to making data visualizations accessible to BLV people, and are recommended by transcription guidelines for the production of accessible charts, diagrams, and maps, where spatial relationships are important~\cite{BANA2010guidelines,rowell2003world}. However, they are time-consuming to produce and costly to update, as any change requires producing a new graphic, making them unsuitable for interactive data exploration. Additionally, He \etal~\cite{He2026} designed 3D printed tactile charts but found that existing design guidelines were inadequate. Tools like Tactile Vega-Lite~\cite{Chen2025} exist to address production of tactile charts, but the resulting graphics remain static. 

Researchers have explored alternative formats. This includes descriptions of charts, where researchers have investigated what makes an effective description~\cite{jung2021communicating,Lundgard2022} and how to automatically generate them~\cite{farahani2023automatic,HoqueEtAl2022chart,alam2023seechart}. However, these do not support spatial exploration. Another approach is using sonification to convey data, which can be used alongside or as a replacement for data visualizations~\cite{Kramer2010}. Recent work has explored combining sonification with speech~\cite{holloway2022infosonics,siu2022supporting}, natural language queries~\cite{SharifEtAlCHI2022}, and tactile graphics~\cite{Ramoa2025}. Other systems support BLV data analysis through screen reader navigation of chart and data structure~\cite{Zong2022,Thompson2023,Elavsky2023}, and authoring environments that combine visualization, sonification, and textual representations~\cite{Zong2024}. However, none of these approaches offer interactive tactile exploration of data visualization through touch.

\subsection{Refreshable Tactile Displays}
Refreshable tactile displays (RTDs), pin-based devices that render tactile graphics, represent a promising option for supporting BLV users in consuming data visualizations. Consisting of pins raised and lowered using electromechanical actuators~\cite{Yang2021}, RTDs offer an advantage over traditional printed tactile graphics: they can display new graphics in seconds. More affordable devices have recently entered the market, including the Dot Pad~\cite{DotInc} (2,400 pins, 60$\times$40, under \$5,000 USD), Monarch~\cite{APH} (3,840 pins, 96$\times$40), and the Graphiti~\cite{Orbit} (2,400 pins, 60$\times$40). Research prototypes such as MagnePins~\cite{Smiley2025} are also exploring affordable, open-source alternatives. While RTDs remain expensive up-front, there is significant value in their interactivity and refreshability.

Research using RTDs has mostly focused on graphics spanning art~\cite{Gyoshev2018}, illustrations~\cite{Kim2019,Namdev2015,Park2016}, accessible maps for navigation~\cite{brayda2019refreshable,Motoyoshi2018,Zeng2015}, and automatic translation of graphics using AI~\cite{Kim2025}. Recent work has begun exploring more interactive and dynamic content on RTDs, including animations~\cite{Holloway2022} and sports visualizations~\cite{Ohshima2021football,Kim2024}. Jiao \etal~\cite{Jiao2025} devised tactile data comics, a step-by-step presentation of tactile graphics with verbal narration on RTDs for educational content. 

Relatively little work has explored how RTDs can support data visualization. Elavsky~\cite{Elavsky2023} co-designed tools for exploring set diagrams, and Holloway \etal~\cite{holloway2024refreshable} surveyed stakeholder perspectives on how RTDs could improve access to data visualization. Seo \etal~\cite{Seo2024,Seo2024b} created MAIDR, which integrates Braille Unicode characters in single-line RTDs for data visualization, though it lacks support for 2D charts. Tactually encoding charts on RTDs remains an open challenge: visual encoding standards do not directly translate to the tactile dimension~\cite{Marriott2026,Khalaila2026,Khalaila2025}, and whether standards for printed tactile charts~\cite{Chen2025,He2026} transfer to RTDs is an open question. 

Chundury \etal~\cite{Chundury2026} compared an RTD with two other accessible data systems: audio-only (screen reader) and audio-tactile (sonification), finding that each modality had limitations in isolation, and recommended combining conversational speech with tactile representation. In our previous work~\cite{Reinders2024}, we used Wizard-of-Oz (WOz) to explore how BLV users interact with data visualization using a system that combined RTDs with a conversational agent, finding preference for multimodal interaction combining speech with touch. The present work builds on these findings by co-designing a fully instantiated system.

\subsection{Conversational Agents and Multimodal Systems}
Advances in Large Language Models (LLMs) have enabled conversational agents that can answer questions about data visualization~\cite{Zhao2025,kavaz2023}. VizAbility~\cite{GorniakEtAl2024Vizibility} allows users to query visual data trends using natural language, and Kim \etal~\cite{Kim2023} mapped queries from BLV users exploring charts to analytical task taxonomies. Choe \etal~\cite{Choe2025} employed an LLM with visual charts; sighted users with higher data literacy engaged more with the charts and utilized the LLM as a specialist, whereas those with lower literacy relied more on the agent. Seo \etal~\cite{Seo2024b} added an LLM to MAIDR~\cite{Seo2024}, and found that BLV users developed strategies to verify LLM-mediated responses. Combining speech with tactile representations has been explored with interactive 3D printed models~\cite{Reinders2020,Reinders2023,Reinders2025} and deictic querying on tactile maps~\cite{ManzoniEtAl2024MapIO}, and our previous WOz work~\cite{Reinders2024} began to explore this for data visualization on RTDs. However, no system exists that tightly integrates multimodal touch input, speech, and tactile feedback for interactive data exploration.

\section{Graphy: A Conversational Tactile Data Interface}
\label{sec:graphy}
\name\ is a CTDI that integrates touch, speech, and tactile feedback for accessible data visualization. To our knowledge, it is the first to combine a multi-line RTD with an LLM-powered conversational agent, supporting line charts (single- and multi-series), bar and stacked bar charts, and scatterplots. \name\ is characterized by:

\bipstart{Scaffolded chart exploration} %(steps 1--4)} 
Charts are split into interactive layers, each narrated by the agent and rendered on the RTD, so users can explore and build their understanding through touch and speech together: one component at a time. Users can skip layers and navigate back and forth at their own pace.

\bipstart{Touch-driven data inspection} %(steps 3, 5, 7)} 
Users explore charts freely by touch, selecting and inspecting data points through double-tap gestures or stepping through values using the RTD's buttons. Each selection is grounded across three feedback channels: an audio label, a Braille label, and a tactile highlight that marks the point on the RTD.

\bipstart{Conversational data analysis} %(steps 6--7)} 
Users can query the agent for trends, calculations, and comparisons, fusing touch and speech by referencing gestured data points directly in their requests (\eg\ selecting two points and asking, \textit{``What is the difference between these two?''}). Responses from the agent are segmented into navigable one-sentence chunks, each synchronized with animated tactile highlights that draw attention to the referenced data on the RTD, allowing users to pace the agent's response alongside their own exploration. 

\bipstart{Filtering for focus} %(step 7)} 
Users can isolate individual data series through voice commands to the agent, reducing chart complexity for more targeted exploration and interaction.

  \begin{figure*}[!htb]
      \centering
      \begin{overpic}[width=0.13\textwidth]{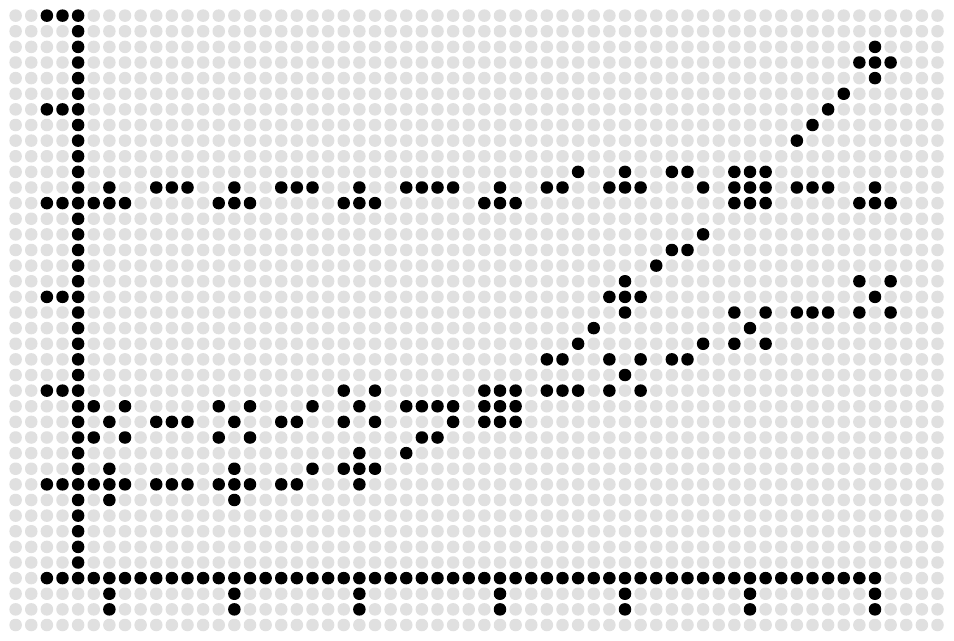}
          \put(40,55){\textbf{(a)}}
      \end{overpic}
      \hfill
      \begin{overpic}[width=0.13\textwidth]{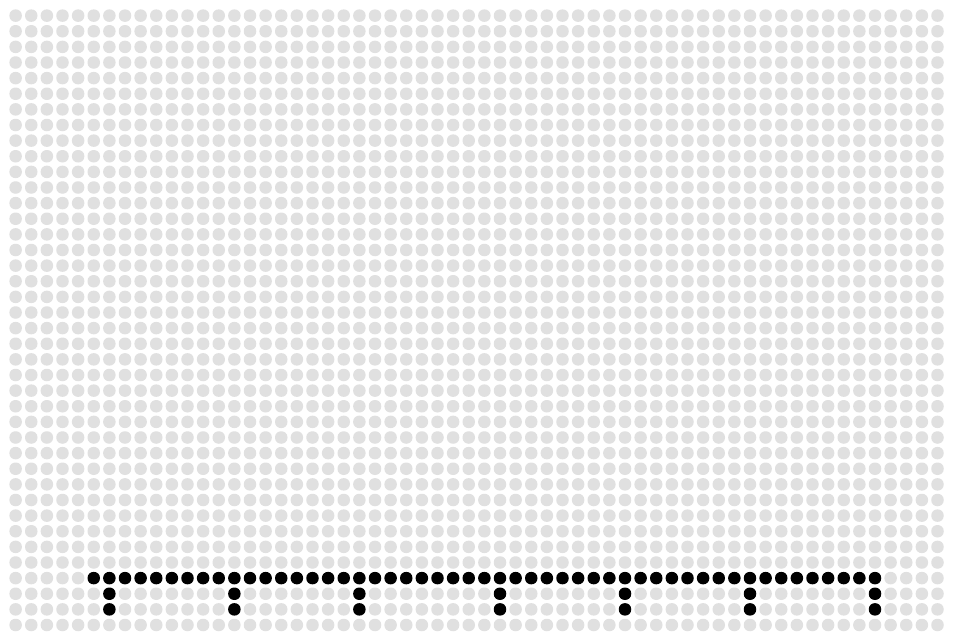}
          \put(40,55){\textbf{(b)}}
      \end{overpic}
      \hfill
      \begin{overpic}[width=0.13\textwidth]{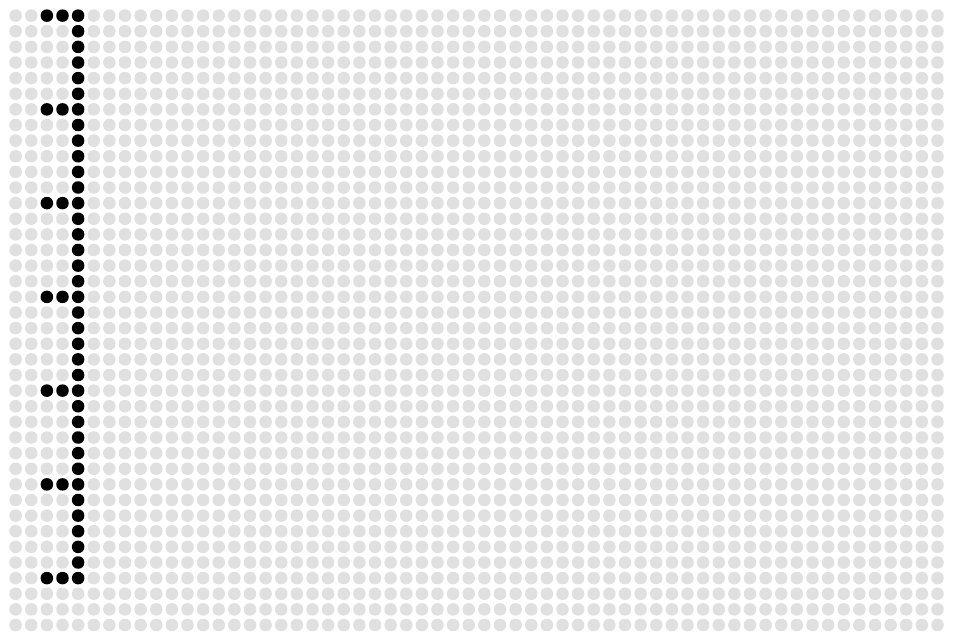}
          \put(40,55){\textbf{(c)}}
      \end{overpic}
      \hfill
      \begin{overpic}[width=0.13\textwidth]{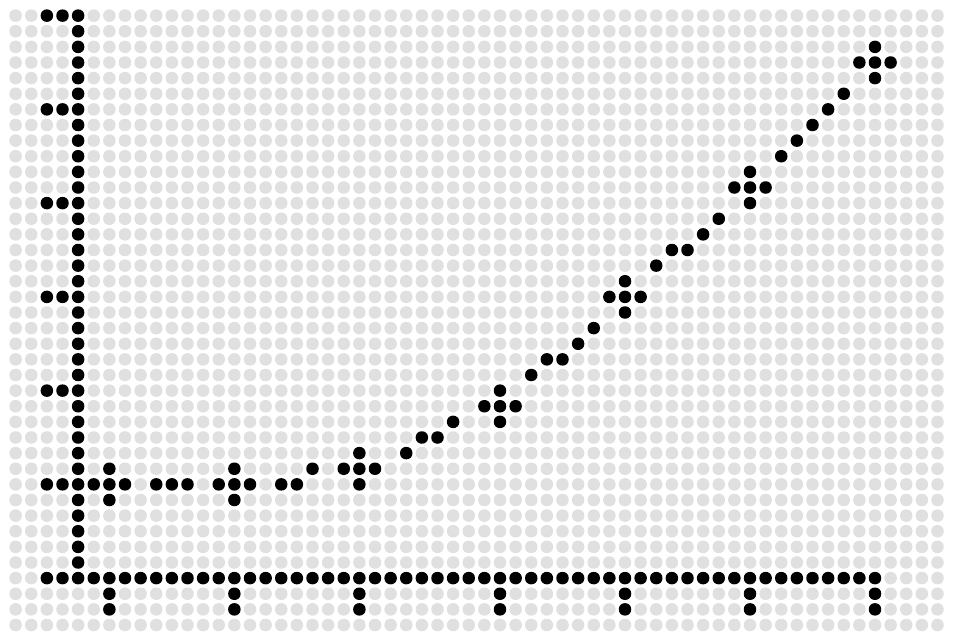}
          \put(40,55){\textbf{(d)}}
      \end{overpic}
      \hfill
      \begin{overpic}[width=0.13\textwidth]{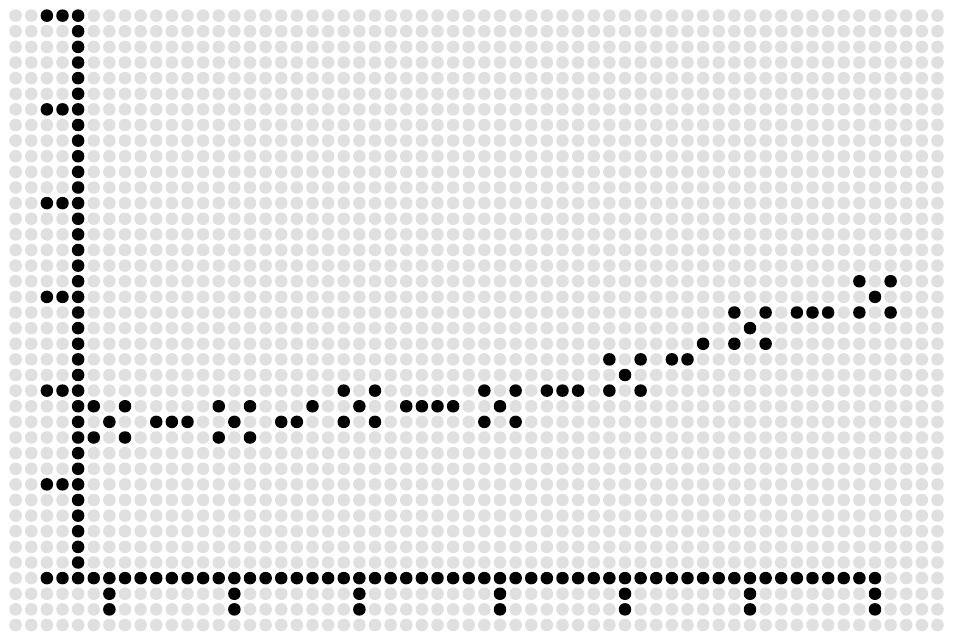}
          \put(40,55){\textbf{(e)}}
      \end{overpic}
      \hfill
      \begin{overpic}[width=0.13\textwidth]{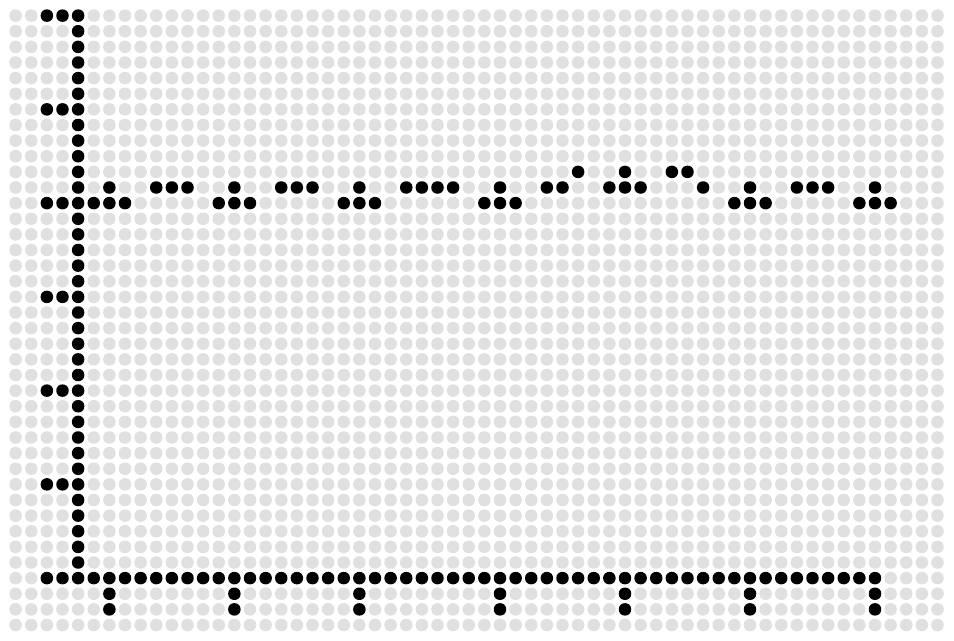}
          \put(40,55){\textbf{(f)}}
      \end{overpic}
      \hfill
      \begin{overpic}[width=0.13\textwidth]{figures/overview/overview_06_summary.pdf}
          \put(40,55){\textbf{(g)}}
      \end{overpic}
      \caption{The layered presentation introducing a multi-series line chart on the RTD. Each layer is presented with a spoken description before the user advances to the next. (a)~title layer with full chart [step 1]; (b)~x-axis layer; (c)~y-axis layer; (d)~series one (Memory) [step 2]; (e)~series two (Storage); (f)~series three (GPU); (g)~summary layer with full chart [step 5].} \label{fig:overview}
      \vspace*{-4mm}
  \end{figure*}

\begin{figure}[h]
      \centering
      \includegraphics[width=0.5\columnwidth]{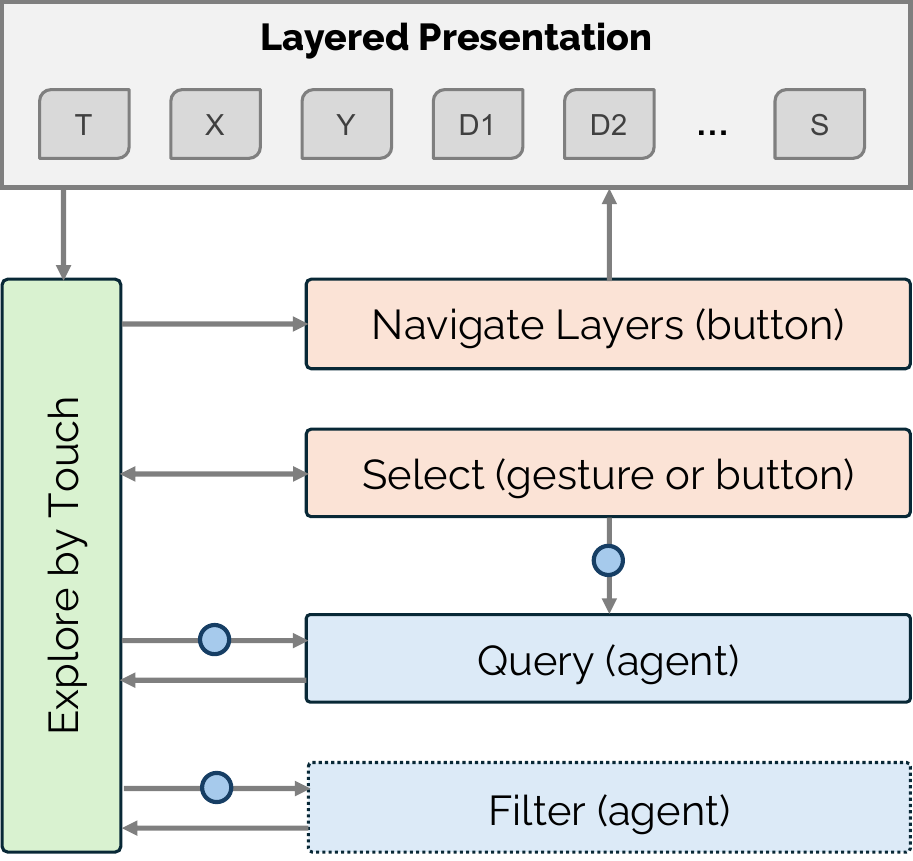}   
      \caption{Interaction flow with \name. Touch is the central activity within the layered presentation (T:~title, X:~x-axis, Y:~y-axis, D1...:~data series, S:~summary); from it, users navigate layers, select data points by gesture or button stepping, query the agent, or filter series. Circles denote agent invocation; the dashed border marks researcher-mediated filtering.}
      \label{fig:interaction}
      \vspace*{-5mm}
  \end{figure}

Together, these capabilities support a fluid interplay of touch and speech. The following scenario, drawn from WS4 (\S~\ref{sec:freeform}), illustrates these in practice, combining tactile exploration with conversational engagement to explore a data visualization. The interaction flow is shown in Fig.~\ref{fig:interaction}, and the manner in which they were co-designed is detailed in \S~\ref{sec:findings}:

\vspace{3mm}
\noindent
\begin{enumerate}
    \myitem D1 holds the push-to-talk button and asks {\bem ``Load the computer component data line chart.''} A multi-series line chart loads on the RTD. The full chart is visible and the layered presentation begins, announcing \textit{``Memory, Storage, and GPU price multi-series line chart, February 2025 to February 2026.''}

    \myitem They begin tracing the line chart, then advance through the x-axis, y-axis, and series one (Memory) presentation layers using the pan right button. Each layer is rendered in isolation on the RTD, and D1 explores by touch as \name\ narrates each component, \eg\ \textit{``Plotted using the plus symbol, Memory prices rose sharply, with two major surges across the period.''}
    
   \myitem In the Memory layer, D1 double-taps the first data point with their left index finger; \name\ produces a static bounding box around the value and announces \textit{``February 2025, Memory, \$190.''} They confirm the selection by touching the box, and proceed to double-tap the last data point with their right index finger; accordingly, a bounding box is rendered and an audio label played.

   \myitem D1 asks {\bem ``What was the average price of Memory between these points?''}; \name\ responds with synchronized speech and animated highlighting on the RTD: \textit{``Your left hand selected Feb 2025, where the price of Memory was \$190, and your right hand selected Feb 2026, where it was \$1100. The average price of Memory between the periods was \$503.''} D1 traces the Memory line between the points to verify the agent's response. 
   
   \myitem D1 advances through the series two (Storage) and three (GPU) layers using the pan right button, reaching the summary layer with the full chart. \name\ announces \textit{``Storage rose steadily, GPU stayed flat, and Memory surged dramatically. Memory intersects with Storage at \$360 in Aug 2025, and meets GPU at \$850 in Dec 2025, before pulling sharply away from both.''}
   
   \myitem D1 asks {\bem ``Filter everything apart from Storage.''} The other series are removed. D1 steps through Storage's data using the buttons, each confirmed with audio and a tactile highlight. They ask {\bem ``What is the trend of Storage?''}; \name\ responds \textit{``Storage prices showed steady growth from \$330 in Feb 2025 to \$600 in Feb 2026...''} D1 traces the Storage line to verify the agent's response. 
\end{enumerate}

\bpstart{Implementation}
\name\ runs on a host computer that drives a Dot Pad RTD (the most affordable multi-line RTD available) over USB serial, and tracks a user's index fingers via an Ultraleap Leap Motion Controller mounted overhead. Speech interaction is invoked by a Picovoice wake word or a push-to-talk button; Google Cloud Speech-to-Text and Text-to-Speech handle transcription and synthesis. Charts are rendered using our Vega-Lite-to-RTD renderer. The agent, built using ChatGPT and LangChain, fuses touch context with queries and answers questions by running pandas code over a chart's dataset, such that values are computed programmatically rather than produced by the LLM. Full architectural detail is in~\cite{Reinders2026}, and open-source code is available.\footnote{\url{https://github.com/accessible-data-vis/feelogue}}

\section{Our Co-Design Method}
\label{sec:method}
We co-designed \name\ with three BLV co-designers. Our aim was to generate design knowledge and deliver a fully functioning implementation through iterative co-design, not to formally evaluate the system. Following Thompson \etal's~\cite{Thompson2023} `\textit{parallel co-design}', we conducted one-on-one rather than group sessions, obtaining unique design insights and personal preferences. Across four rounds of workshops (twelve individual sessions over eight months), we prioritized longitudinal engagement over single-session studies with more participants, consistent with co-design practice in accessible visualization~\cite{Zong2022,Elavsky2023,Chundury2024,Seo2024,Thompson2023}.

\subsection{Co-Designers}
We recruited three co-designers\footnote{Following the co-design process, co-designers were invited to be listed as co-authors in recognition of their contributions\cite{Mankoff2010}.} from our lab-managed participant pool, each with prior design experience from unrelated research projects. The project was approved by our institutional ethics committee (ID: 37871), and all co-designers provided informed consent. Their tactile-graphics experience ranged from some (D3) to substantial (D1, D2), and all had used RTDs, voice assistants (Alexa, Google Assistant, Siri), and the LLM-based agent ChatGPT (Tab.~\ref{tab:designers}). All were comfortable working with data for basic numerical tasks; D1 and D2 were confident in performing more complex tasks (\eg\ statistical analysis and interpreting trends), whereas D3 reported limited confidence beyond interpreting line charts and was neutral on more advanced data analysis. 

\begin{table}[t]
\setlength{\tabcolsep}{5pt}
\caption{Co-designer demographics. Level of blindness, onset of vision loss, and tactile graphics (TG) and RTD experience are self-rated.\label{tab:designers}}
%%(None, Limited, Some, Substantial)
\small
\centering
\begin{tabular}{lccccc}
\toprule
\textbf{D\#} & \textbf{Age} & \textbf{Blindness} & \textbf{Onset} & \textbf{TG Experience} & \textbf{RTD Experience}\\
\midrule
D1 & 24 & Legally Blind & Congenital & Substantial & Some\\
D2 & 27 & Totally Blind & Congenital & Substantial & Substantial\\
D3 & 55 & Totally Blind & Congenital & Some & Some\\
\bottomrule
\end{tabular}
\end{table}

\subsection{Workshops}
\begin{figure*}[t!]
      \centering
      \begin{overpic}[width=0.16\textwidth]{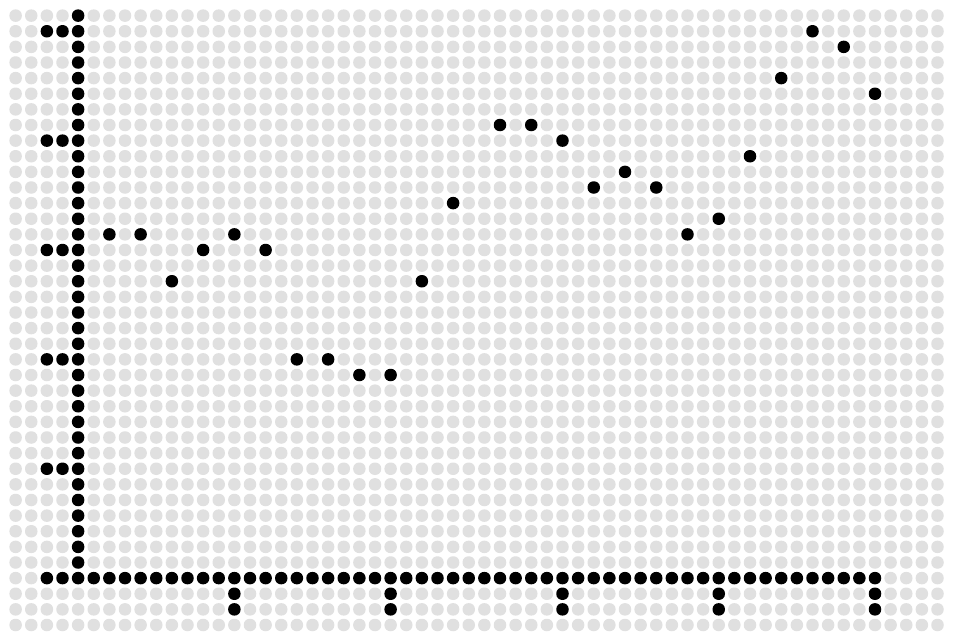}
          \put(40,57){\colorbox{white}{\textbf{(a)}}}
      \end{overpic}
      \hfill
      \begin{overpic}[width=0.16\textwidth]{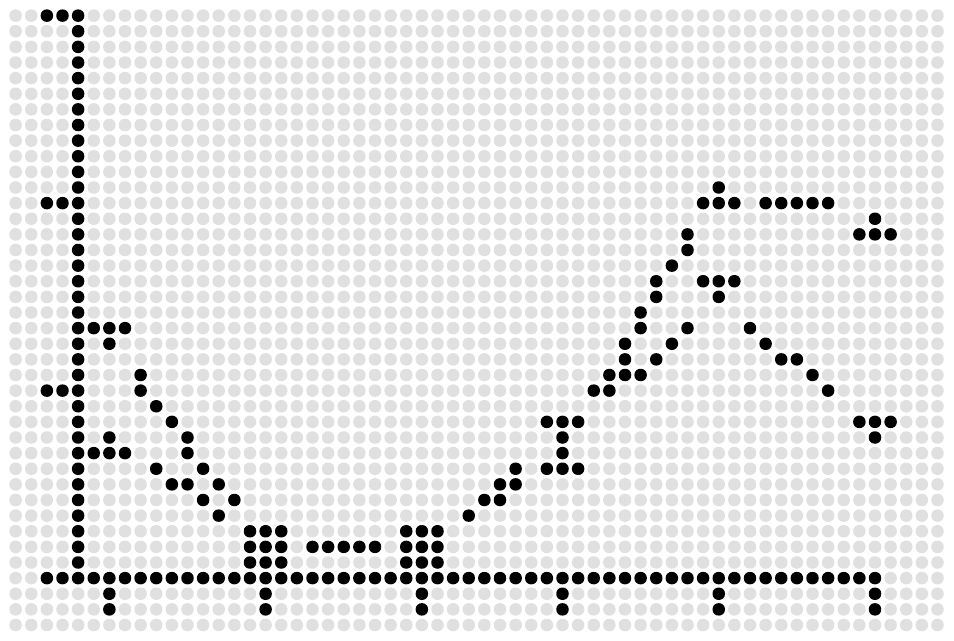}
          \put(40,57){\colorbox{white}{\textbf{(b)}}}
      \end{overpic}
      \hfill
      \begin{overpic}[width=0.16\textwidth]{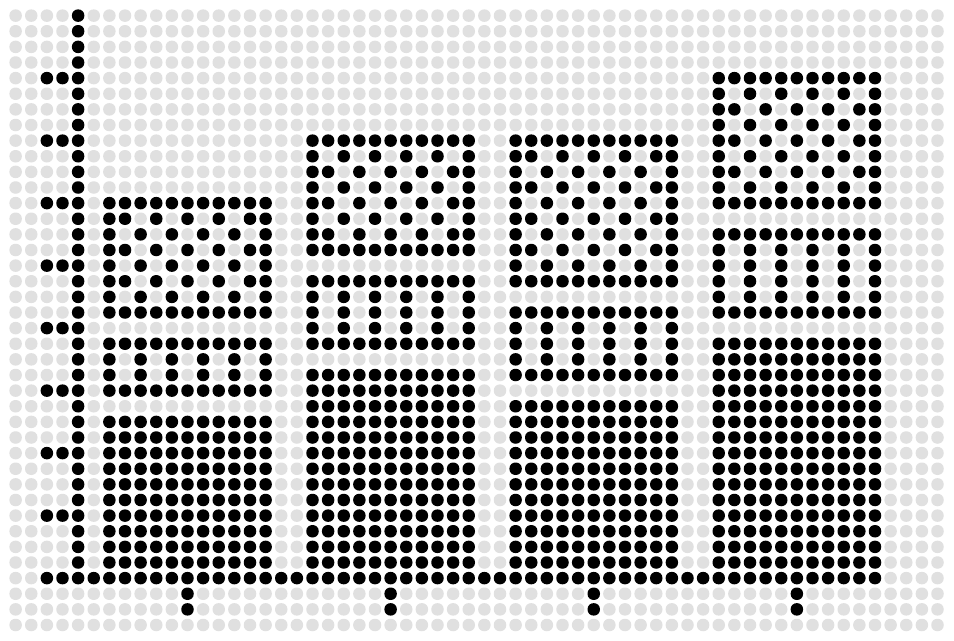}
          \put(40,57){\colorbox{white}{\textbf{(c)}}}
      \end{overpic}
      \hfill
      \begin{overpic}[width=0.16\textwidth]{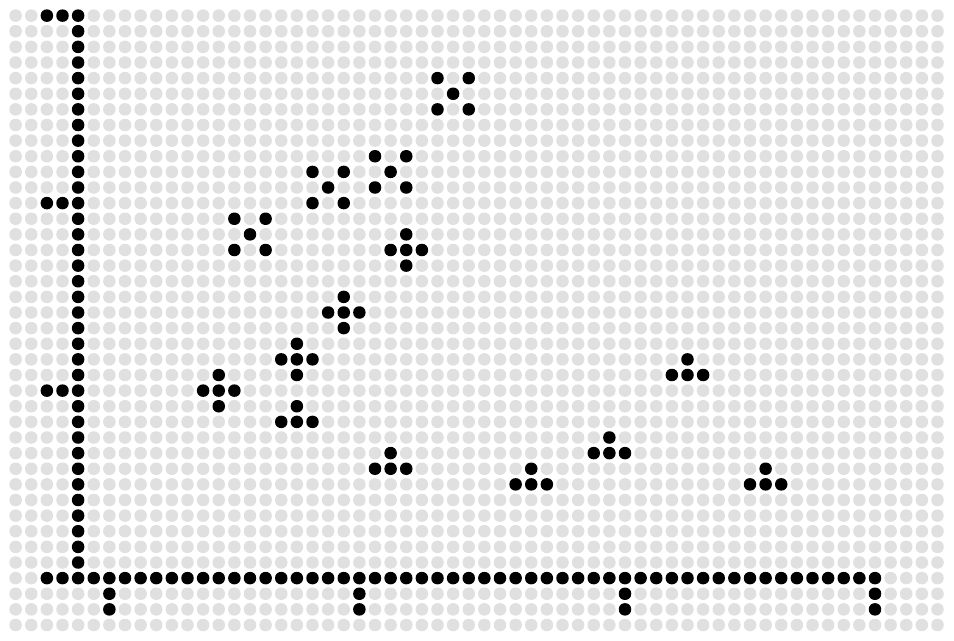}
          \put(40,57){\colorbox{white}{\textbf{(d)}}}
      \end{overpic}
      \vspace*{-2mm}
    \caption{Renderings of the data sets used across workshops. (a)~line chart of annual rainfall (WS1), (b)~multi-series line chart of interest rates (WS3), (c)~stacked bar chart of quarterly product revenue (WS3), (d)~scatterplot of vehicle efficiency and power.}
    \label{fig:charts}
    \vspace*{-5mm}
\end{figure*}

Co-designers participated in all four workshops and were compensated after each session with a gift card, pro-rated at \$50 AUD per hour. Successive rounds were spaced two months (WS1--WS2), three months (WS2--WS3), and one week (WS3--WS4) apart. These intervals reflected the time required to discuss, implement and stabilize design changes between rounds. Datasets were drawn from real-world sources, and included annual rainfall, interest rates, quarterly product sales and vehicle efficiency data (Fig.~\ref{fig:charts}). Across workshops, co-designers were asked to use the `\textit{think-aloud}' method~\cite{Wobbrock2009}, verbally describing intended interactions, expectations, and reasoning. 

We used design probes to ground design discussions in interactive artifacts rather than abstract feature descriptions. Probes included both researcher-proposed designs informed by our prior work~\cite{Reinders2024} and alternatives proposed by co-designers, organized around five areas: 
\begin{itemize}
    \item \textbf{Orientation \& Overview:} How \name\ introduces and orients users to new charts.
    \myitem \textbf{Chart Tactile Encoding:} How data is represented tactually across chart types on the RTD.
    \myitem \textbf{Navigation, Selection \& Interaction}: How users want to navigate, select, and interact with the data.
    \myitem \textbf{Tactile Feedback}: How the system provides tactile feedback to confirm user actions and draw attention to agent-referenced elements on the RTD.
    \myitem \textbf{Agent Interaction}: How users engage with \name's conversational agent and the role it plays. 
\end{itemize}

These probes were integrated into workshops WS1--WS3, each of which consisted of three parts:
\begin{itemize}
    \item \textbf{(Re)-Familiarization \& Changes:} In WS1, co-designers were introduced to the system's functionality and asked to try each interaction. In WS2, they were re-introduced to the system and shown all changes made since WS1. In WS3, co-designers were instead asked to recall how each interaction worked from memory, with researchers filling in gaps as needed, before being shown the changes made since WS2. 
 
    \myitem \textbf{Hands-On Activity}: Co-designers explored datasets using \name, with each round introducing new chart types: single-series line charts (WS1), bar charts (WS2), and multi-series line charts, stacked bar charts, and scatterplots (WS3). Activities progressed from independent exploration (WS1) to hands-on engagement structured around design probes (WS2--WS3).
        
    \myitem \textbf{Design Feedback \& Ideation:} Open discussions, forming the bulk of each session, used design probes to prompt reflection across tactile, conversational, and multimodal aspects of \name. Probes were initially based on design challenges identified by the research team (WS1) and increasingly shaped by co-designer feedback (WS2--WS3).
\end{itemize}

WS4 followed a different structure, aiming to observe how memorable the co-designed system was through independent use with co-designers' own data. Co-designers were asked to form a summary of their data that they could share with a friend. Afterward, they completed a NASA-TLX workload assessment---used as a reflective prompt rather than an evaluative measure---and a question assessing perceived naturalness, followed by a semi-structured reflection on how the system had evolved throughout the co-design process. 

\subsection{Analysis \& Iteration}
All sessions were video and audio-recorded and transcribed. WS1 and WS2 were attended by two researchers, and WS3 and WS4 by one. After each round, the attending researchers reviewed the transcripts, categorizing comments into five areas: new feature requests, system bugs, system improvements, and positive and negative experiences. Proposed changes were discussed with the broader research team and implemented before the next workshop round.

Our findings (\S~\ref{sec:findings}) were derived from a design synthesis across workshop rounds, tracing recurring patterns and changes in preferences and workflows over time. For WS4, we also report workload and naturalness ratings, along with co-designers' reflections.

\section{Co-Designing \name: Design Findings}
\label{sec:findings}
 The co-design process began with an initial prototype (v1) informed by our WOz findings~\cite{Reinders2024}. It supported the display of single-series line charts with basic touch and speech interaction, including: an LLM-generated audio overview upon chart loading, double-tap gesture selection with audio and Braille labeling, free touch during speech for deictic references to the agent, wake-word agent activation, and single-pin blinking for highlights (selected and agent-referenced values). The agent supported data analysis backed by Python data libraries, including trend analysis, comparisons, and statistical calculations.

In this section, we trace how \name\ transformed from this v1 prototype into the system described in \S~\ref{sec:graphy}. In doing so, we expand the design space for CTDIs. We present findings across five design areas, followed by observations from free-form use and retrospective reflections. Within each area, we describe how co-designers shaped the design through feedback and ideation across the workshops (Tab.~\ref{tab:features}).

\begin{table}[t]
    \caption[Features across the co-design process]{Features across the co-design process: 
    (\protect\tikz[baseline=-0.5ex]{\draw[line width=0.6pt] (0,0) circle(2pt);})~suggested, 
    (\protect\tikz[baseline=-0.5ex]{\draw[line width=1pt, dashed] (0,0) -- (3mm,0);})~not yet implemented,
    (\protect\tikz[baseline=-0.5ex]{\fill (0,0) circle (2pt);})~implemented/refined, 
    (\protect\tikz[baseline=-0.5ex]{\draw[line width=1pt] (0,0) -- (3mm,0);})~carried forward.\label{tab:features}      
    \vspace*{-2mm}
}
    \centering
    \setlength{\tabcolsep}{5pt}
    \small
    \begin{tabular}{lYYYYY}
    \toprule
    \textbf{Feature} & \textbf{v1} & \textbf{WS1} & \textbf{WS2} & \textbf{WS3} & \textbf{WS4}\hspace*{-3mm}\\
    \midrule
    \multicolumn{5}{l}{\textit{\textbf{Orientation \& Overview}}}\\
    \hspace{1em}LLM narrated overview & \patA\\
    \hspace{1em}Layered overview & \patL\\
    \midrule
    \multicolumn{5}{l}{\textit{\textbf{Chart Tactile Encoding}}}\\
    \hspace{1em}Single-pin data points & \patA\\
    \hspace{1em}Connecting lines & \patL\\
    \hspace{1em}Point symbols & \patI\\
    \hspace{1em}Fill textures & \patJ\\
    \midrule
    \multicolumn{5}{l}{\textit{\textbf{Navigation, Selection \& Interaction}}}\\
    \hspace{1em}Double-tap gesture & \patC\\
    \hspace{1em}Free touch during speech & \patA\\
    \hspace{1em}Sequential interaction & \patE\\
    \hspace{1em}Button stepping & \patM\\
    \hspace{1em}Data filtering & \patJ\\
    \midrule
    \multicolumn{5}{l}{\textit{\textbf{Tactile Feedback}}}\\
    \hspace{1em}Single-pin blinking & \patA\\
    \hspace{1em}Static highlights & \patI\\
    \hspace{1em}Animated highlights & \patJ\\
    \midrule
    \multicolumn{5}{l}{\textit{\textbf{Agent Interaction}}}\\
    \hspace{1em}Wake-word activation & \patB\\
    \hspace{1em}Push-to-talk activation & \patK\\
    \hspace{1em}Deictic queries & \patO\\
    \hspace{1em}Response format & \patH\\
    \hspace{1em}Response segmentation & \patK\\
    \hspace{1em}Automatic follow-ups & \patK\\
    \hspace{1em}Data analysis operations & \patB\\
    \bottomrule
    \vspace*{-8mm}
    \end{tabular}
\end{table}

\subsection{Orientation \& Overview}
Sighted users can orient themselves in a new chart by instantly surveying its shape, peaks, and trends. In contrast, when BLV users read tactile charts, touch exploration is a sequential experience: they move their hands across the graphic to incrementally build a mental model, often focusing on local features before understanding the whole. The v1 prototype attempted to support this by automatically presenting an overview on chart load: the chart title, x-axis range, y-axis range, and a brief interpretive summary of the data narrated as a single continuous description, akin to alt text or a transcriber's note. While exploring the chart through touch simultaneously, co-designers found it difficult to follow due to its length, and often replayed it. 

\bpstart{Layered overview}
D1 proposed replacing the single narration with a layered, sequential introduction of chart components (WS1). We implemented this: each component -- title, x-axis, y-axis, data, and interpretive summary -- was introduced one at a time, rendered cumulatively on the RTD, with button press to advance (WS2). D3 described this as follows: \textit{``It's one thing at a time, and you only move on to the next complication when you're comfortable$\ldots$ it's a scaffolding thing, we just start gently and then build up.''}

\bpstart{Shaping overview behavior}
The v1 overview supported touch exploration but not interaction; co-designers insisted on being able to select data points and query the agent during the overview, to support active sensemaking rather than passive listening (WS2). D3 spoke of the overview as replacing the role of a sighted guide: \textit{``The [need for a] human is replaced by what it [the system] says.''} Co-designers also agreed that experienced users could skip the overview (D2: \textit{``If it was a data set I had already looked at a lot$\ldots$ I would want an easy way to skip the thing''}). D2 distinguished between structural and interpretive content, wanting to know \textit{``the domain and the range and what the symbols are,''} but not the system's interpretive summary: \textit{``I'll interpret it.''} Despite this, the included components and their ordering (title $\rightarrow$ x-axis $\rightarrow$ y-axis $\rightarrow$ data $\rightarrow$ summary) were consistent across co-designers.

\bpstart{Isolation for focus}
As more complex chart types were introduced, \eg\ multi-series charts (WS3), the overview became more valuable (D1: \textit{``[It was] more useful now than it was with a single line''}), but multiple overlapping series make cumulative layering tactually cluttered. The co-designers suggested presenting components in isolation: showing the x-axis alone, then the y-axis alone, then each data series individually (with the axes for spatial context) to improve focus (Fig.~\ref{fig:overview}). As D2 explained, \textit{``Show only the first one [series], then show only [the second]$\ldots$ you build the mental model of each line in isolation and then you put that together''.} The co-designers felt this approach could apply to all chart types and complexity levels. The summary layer returned the complete chart with all components visible, allowing co-designers to explore the full chart while listening to the summary.

\subsection{Chart Tactile Encoding on RTDs}
When sighted users read a chart, they leverage multiple visual channels (\eg\ position, shape, size, color) simultaneously. On the majority of RTDs, tactile output is binary --- each pin is either raised or lowered --- and constrained by low resolution (60$\times$40), requiring encoding decisions that balance tactile readability against clutter. The v1 prototype rendered simple line charts using single raised pins for data points. With densely sampled data, co-designers could trace without explicit connecting lines, but noted that sparser data would require guessing the direction of change between points. The prototype also lacked symbols to distinguish series or encodings for other chart types.

\begin{figure}[ht]
    %%\centering
    \vspace*{-2mm}\includegraphics[width=\columnwidth]{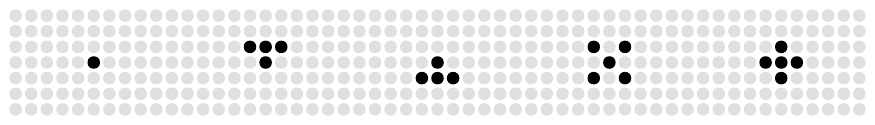}
    \vspace*{-2mm}\hspace*{8mm}(a)\hspace*{13mm}(b)\hspace*{13mm}(c)\hspace*{14mm}(d)\hspace*{14mm}(e)
    \caption{Symbols used to distinguish data points. From left to right: (a)~single dot, (b)~arrow down, (c)~arrow up, (d)~X, (e)~plus. Co-designers ranked the plus as the most tactually distinguishable (WS3).}
    \label{fig:symbols}
    \vspace*{-4mm}
\end{figure}

\begin{figure}[ht]
  \centering
  \begin{overpic}[width=0.275\columnwidth]{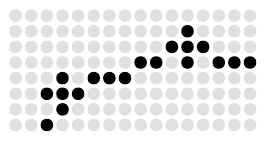}
      \put(10,37){\textbf{(a)}}
  \end{overpic}
  \hfill
  \begin{overpic}[width=0.275\columnwidth]{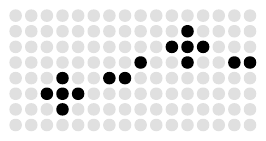}
      \put(10,37){\textbf{(b)}}
  \end{overpic}
  \hfill
  \begin{overpic}[width=0.275\columnwidth]{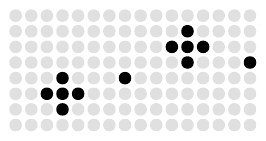}
      \put(10,37){\textbf{(c)}}
  \end{overpic}
  \vspace*{-2mm}
  \caption{Symbol-to-line clearance gaps between two plus symbols connected by a line: (a)~0-pin gap, (b)~1-pin gap, (c)~2-pin gap.}
  \label{fig:clearance}
  \vspace*{-3mm}
\end{figure}

\bpstart{Connecting lines and symbols}
Connecting lines were introduced between data points (WS2). All co-designers responded positively, with D2 describing them as \textit{``filling in the truth$\ldots$ you don't need to guess the direction [of the data].''} However, single-pin markers blended into the line (something we did not anticipate), making it difficult to distinguish data points from the line itself. We introduced point symbols to mark data points, and tested four: arrow down, arrow up, cross, and the plus symbol (Fig.~\ref{fig:symbols}). Co-designers drew on their experience with printed tactile graphics, suggesting familiar symbols (\eg\ circles and 2$\times$2 squares), but these proved harder to distinguish on the RTD. Co-designers ranked the plus as the most readable, with the cross ranked second. The arrows were easy to locate through touch, but ranked last. Co-designers suggested a one-pin clearance gap between symbols and connecting lines to enhance readability (Fig.~\ref{fig:clearance}). D3 said, \textit{``You can [now] determine the symbols. They're nice and separate from everything else,''} while D1 preferred an adjustable setting.

\bpstart{Multi-series line charts}
As multi-series line charts were introduced, new encoding challenges emerged around overlapping and adjacent lines (WS3). For intersections, D3 suggested a 3$\times$3 square symbol. For adjacent lines, line thickness was tested, but co-designers were divided: D2 felt that thicker lines made \textit{``things way more cluttered''}, while D3 described them as \textit{``100 times better.''} The difficulty of reading more complex charts motivated D1 to suggest filtering (\S~\ref{sec:interaction}).

\begin{figure}[ht]
    %%\centering
    \vspace*{-2mm}\includegraphics[width=\columnwidth]{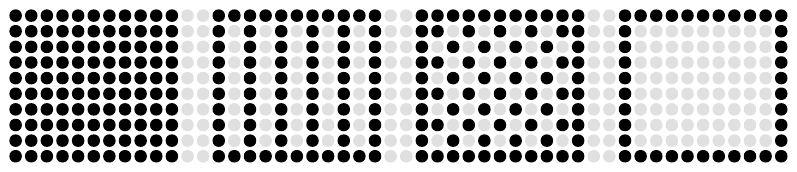}
      \vspace*{-2mm}\hspace*{8mm}(a)\hspace*{20mm}(b)\hspace*{18mm}(c)\hspace*{20mm}(d)
    \caption{Fill textures used to distinguish segments in stacked bar charts: (a)~solid infill, (b)~vertical bars, (c)~checkerboard, (d)~hollow perimeter.}
    \label{fig:textures}
    \vspace*{-3mm}
\end{figure}

\bpstart{Bar charts}
Bar charts were introduced in WS2, and stacked bar charts in WS3. D3 had never encountered a bar chart before the co-design sessions, initially reacting with dislike (\textit{``It's never the way that I [have] ever had data represented for me$\ldots$ yuck''}). By the end of WS3, they were independently reading and comparing stacked segments within a single session. For stacked bar charts, co-designers emphasized that fill textures and clear segment delineation were essential. We explored five textures adapted from tactile graphic guidelines (Fig.~\ref{fig:textures}). One (horizontal bars) was rejected as it could not render in segments smaller than three rows; the remaining four --- solid infill, vertical bars, a checkerboard pattern, and a hollow perimeter --- were all distinguishable by touch without being named. Segments were separated by a one-pin gap with a solid boundary line (D1: \textit{``With textures and a gap you've got two kind of tactile indicators''}), and adjacent bars by a two-pin gap. Co-designers could clearly feel where one bar ended and the next began, but felt it should be configurable. When textures were replaced with a solid fill during a design probe, readability dropped sharply (D1: \textit{``Way less readable, actually not even readable at all''}).

\bpstart{Scatterplots}
Co-designers initially had difficulty interpreting scatterplots, needing to traverse the layered overview more than once to build a conceptual understanding (WS3). They raised concerns about symbol overlap: D2 felt that slight overlap (1 pin) was manageable from a readability perspective, but greater overlap would require jittering or some form of aggregation, while D1 suggested reusing the same intersection symbol from multi-series line charts. 

\subsection{Navigation, Selection \& Interaction}
\label{sec:interaction}
With a tactile chart rendered on an RTD, BLV users can only perceive what is directly under their fingertips: touch is sequential and local, so reaching a data point means first locating it---and risking loss of spatial context along the way. The v1 prototype supported double-tap gestures for selecting data points and retrieving audio labels, and deictic references to the agent through gestures or free touch during speech.

\bpstart{Gesture selection}
The double-tap gesture for selecting data points was well received, with co-designers noting its familiarity from mobile screen readers (WS1). As finger tracker reliability improved across sessions (tuned to co-designers' touch-reading patterns), gestural input became increasingly valued; by WS3, D1 rated it as \name's \textit{``most important feature,''} noting, \textit{``So cool, this actually [makes me feel] like a child on Christmas Day.''} Initially, selecting a data point triggered audio and Braille labels, but co-designers wanted a tactile cue to confirm correct selection, leading to the introduction of tactile highlighting (WS2, \S\ref{sec:highlighting}). All three channels would be activated simultaneously: audio for immediate confirmation, Braille for verification, and a tactile highlight for spatial confirmation.

\bpstart{Sequential interaction}
Co-designers shaped how selection and querying should be conducted, preferring a sequential flow: exploring by touch, selecting values using gestures, confirming them via an audio or Braille label, and then asking the agent questions (WS1). They wanted to verify they had selected the intended point before querying the agent; without that confirmation, they risked asking about the wrong data point. Once a point of interest was found through touch exploration, each step confirmed the previous: \textit{select, confirm, ask}. This contrasted with our WOz study~\cite{Reinders2024}, where participants preferred free touch during speech, which offered no confirmation step between selection and query; we subsequently focused on gestural input.

\bpstart{Stepping through data points}
D1 suggested that the RTD's physical buttons could be used to step through data points, noting that it would \textit{``allow [them] to get to the next value without having to find it''} (WS2). Implemented in a probe (WS3), stepping is analogous to a screen reader, allowing users to move through values one at a time. Each stepped value is confirmed with the same triple feedback as gesture selection: audio and Braille labels, and a tactile highlight. D2 used stepping to improve selection precision and for recovery when gestures overshot, while D3 used the narrated values to guide directional touch exploration, predicting the shape of the data before tracing it by hand, calling it \textit{``voice-guided exploration.''} 

For multi-series and stacked bar charts, stepping order proved task-dependent (D3: \textit{``Am I wanting to look at software only [same segment across bars]? Or quarter per quarter [all segments within a bar]?''}). For scatterplots, D2 noted that users must choose which dimension to traverse: \textit{``You kind of have to select the [dimensions] to fix, which means you're traversing one.''}

\bpstart{Filtering for focus}
As chart complexity increased, co-designers found it difficult to parse overlapping lines for interaction (D2: \textit{``More cluttered$\ldots$ this feels very noisy to me''}). D1 proposed filtering to isolate individual series, describing it as \textit{``critical to reading charts, particularly more complicated plots.''} Filtering was seen as a way to reduce tactile clutter and sharpen focus for touch reading and interaction by narrowing the data points available for gestures and stepping.

\subsection{Tactile Feedback}
\label{sec:highlighting}
Designing effective tactile feedback on RTDs is challenging: feedback must be easy to locate by touch and meaningful. The v1 prototype used single-pin blinking, at an interval similar to a blinking cursor, for confirming gesture selections and agent-referenced highlights. However, co-designers found this feedback too subtle, and often missed it; D1 described the v1 blinking as \textit{``easy to miss unless you were already touching that area.''} This problem was compounded by a hardware limitation: the electromagnetic actuators of the Dot Pad (and similar RTDs) lack sufficient force to raise pins against finger pressure, meaning pins do not reliably actuate when being touched. 

\bpstart{Static highlights} 
We suggested a static bounding box as an alternative way of drawing attention (WS2), which co-designers found easy to distinguish and locate. D2 described their approach as \textit{``You familiarize with the pattern [shape], and then scan to find it.''} A crosshair pattern was also suggested (D2: \textit{``X marks the spot''}), but was too large and similar to the cross symbol. Co-designers found that the bounding box also served as a spatial anchor, allowing them to return to a previous point after removing their hands (D3: \textit{``I want to know where I am$\ldots$ then I go across the chart [exploring] and find [return to] the box''}). Each finger can maintain one highlight at a time, allowing co-designers to mark and compare two points simultaneously. D3 also suggested an isolation mode, which would temporarily lower and hide nearby chart elements to increase contrast around highlighted points (WS2). 

\bpstart{Animated highlights and feedback grammar}
Both user- and agent-initiated highlights initially used the same static bounding box (WS3). This became problematic because gesture selection, stepping, and agent references rendered the same pattern: co-designers had to infer whether a highlight was in response to their own action or something the agent was drawing attention to. All co-designers arrived at the same solution: static highlights work for user selections because they already know where they selected, whereas agent-initiated highlights require animation to draw attention. Co-designers explored animating the data point symbols, but found them harder to interpret than the bounding box: they had to first locate the animation, then determine which symbol was animating and what it represented, adding cognitive overhead.

D1 refined this into a feedback grammar: gesture selection = static box; stepping = animation that settles to static; agent reference = animated box. Co-designers transferred this grammar to bar charts (D3: \textit{``For consistency, let's still use animation for when we ask questions''}), though they proposed using the bar's infill textures rather than a bounding box: gesture selection would remove infill, stepping would animate and then settle, and agent references would continuously animate. Co-designers agreed that all highlights should persist until dismissed.

\subsection{Agent Interaction}  
Combining a conversational agent with an RTD raises the question of how it will be used when users can already feel the data. The v1 agent was invoked via a wake word (\textit{``Hey Graphy''}), with responses generated without constraints on format or length. Co-designers found the wake word unreliable (WS1), and indicated it was difficult to determine when the agent was listening and when it was processing their request. 

\bpstart{Agent invocation}
All three co-designers suggested a push-to-talk button (WS1). Although this requires briefly lifting a hand off the chart, which can cause a momentary loss of spatial context, they felt that deterministic control outweighed this cost (D2: \textit{``Walkie-talkie with a robot''}). The wake word was retained for situations requiring both hands on the chart, but seldom used. Co-designers also replaced v1's uniform start/stop listening earcon tone (based on Siri) with distinct ascending and descending tones (based on Google Assistant): \textit{``Low to high, you speak, high to low closes the loop''} (D3) --- valued as confirmation that their input had registered, even with push-to-talk. 

D1 expected the agent to support automatic follow-ups when a query lacked specificity, prompting the user for clarification by using the same earcon cues rather than requiring re-invocation (\textit{``That's cue to give more information$\ldots$ that's what LLMs do''}), noting this would be important for complex analytical queries involving multiple steps.

\bpstart{Role of the agent}
Without being told what the agent could or could not do, co-designers directed queries to the agent that touch alone could not resolve, including: calculations and comparisons (\eg\ \textit{``Is this point lower than the average?''}), trend analysis (\eg\ \textit{``Describe the overall trend''}), visualization literacy (\eg\ \textit{``How does a scatterplot work?''}), and operations (\eg\ \textit{``Filter everything''}). The complexity of queries grew across workshops, from basic value retrieval (WS1) to queries about trends and queries with deictic references (WS2), and requests for comparative analysis (WS3). D2 also expected the agent to explain its capabilities to improve discoverability: \textit{``What does this button do$\ldots$ what can I ask you?''}

\bpstart{Agent responses}
Co-designers noted the chart already provides spatial context through touch. Hence, they wanted concise, answer-first responses (\eg\ \textit{``\$120 in 2010''} rather than \textit{``The stock price in 2010 was \$120''}), \textit{``Not all the fluff''} (D3). However, overly terse responses (\eg\ \textit{``\$120''}) were rejected, as they lacked context to verify correctness (D3: \textit{``What if it misheard me? I think you need to include my parameters in your answer''}). Response length was seen as task-dependent, with D2 likening length to \textit{``skim-reading versus depth-reading.''} They also insisted on consistent structure (D1: \textit{``The information has to be in the same format''}), which requires precise prompting of the LLM.

\bpstart{Touch context}
In deictic queries, users fuse touch and speech by referencing selected data points directly in their query. Co-designers wanted the agent to acknowledge the touch context before answering, so they could confirm that the correct point(s) were selected. Without this, D1 noted \textit{``I just assume [that the agent] has picked a random spot, and I don't know where.''} D3 similarly valued this as confirmation the agent had correctly interpreted their interaction, preferring the agent to describe hand positions explicitly (\eg\ \textit{``Your left hand is touching 2010, where the average was X, and your right is touching 2015$\ldots$''}).

\bpstart{Response segmentation}
Co-designers found it difficult to follow long agent responses when engaging in touch exploration, with D3 noting that \textit{``longer responses can be overwhelming''}, as audio and touch compete for attention. D1 and D3 suggested breaking responses into sentence-sized chunks navigable via the RTD's buttons, allowing users to pace responses to match their exploration speed (WS1). Each segment was synchronized with corresponding tactile highlights, linking the agent's spoken response to spatial locations on the chart. D1 described this as a \textit{``game changer for keeping track''} (WS2).

\subsection{Free-Form Use \& Reflection}\label{sec:freeform}
In WS4, co-designers explored a multi-series line chart rendered from data that they had suggested at the end of WS3: computer component pricing~(D1), computer storage format pricing~(D2), and census language data~(D3). Each chart contained three series. With no familiarization component and WS3 sessions held one week prior, we wanted to assess how memorable and intuitive the system was. Tasked with forming a summary of the data that they could share with a friend or colleague, co-designers were free to use \name\ however they chose. The free-form activity lasted between 28 and 39 minutes (average 32).

\subsubsection{Observed Interaction Patterns}
All co-designers followed a similar interaction pattern centered on the layered overview: they navigated back and forth between layers throughout their exploration, selecting and inspecting values via gestures and stepping, querying the agent for what touch could not resolve, and building their understanding of the data within and across layers (Tab.~\ref{tab:ws4}). D2, who had initially described the layered overview as \textit{``gimmicky''} (WS2), had become its strongest advocate by WS4.

There was a clear division of labor: touch was the primary sensemaking channel, gesture/button selection supported precision inspection, and the agent was used for targeted queries such as data retrieval (D3: \textit{``How many Mandarin speakers were there in 1991?''}), trend analysis (D1: \textit{``What is the trend of Memory in this time period?''}), and cross-series comparison (D2: \textit{``When is the cost of all three storage media projected to be equal?''}). Filtering was performed through speech (D1: \textit{``Filter everything apart from Memory''}), functioning as a focusing strategy to reduce chart complexity and allow targeted exploration; filtering commands were mediated by the researcher behind the scenes.

However, there were individual differences between co-designers. D2 was the most Braille-reliant, reading labels after nearly every selection to verify the agent's responses; D3 was the most touch-dominant, with deliberate, repeated tracing and comparing along and across series; and D1 made the most use of filtering, progressively isolating individual series to increase focus, before comparing across them. 

All the co-designers were able to produce substantive summaries of their dataset that went beyond the system's interpretive summary. D1 identified that the price of Memory surged from cheapest to most expensive, noting that \textit{``in October something happened to Memory where it just went ballistic, I wonder what it was$\ldots$ data centers?''}; D2 described SSD and Hard Disk reaching price parity and asked the agent to project when/if all formats would converge; and D3 traced three languages, describing the steady decline of Italian against Mandarin's rise over 30 years, personifying the two series (\textit{``Previously I was more than you, and then we became the same, and then you overtook me''}).

\begin{figure}[h]
      \centering
      \includegraphics[width=\columnwidth]{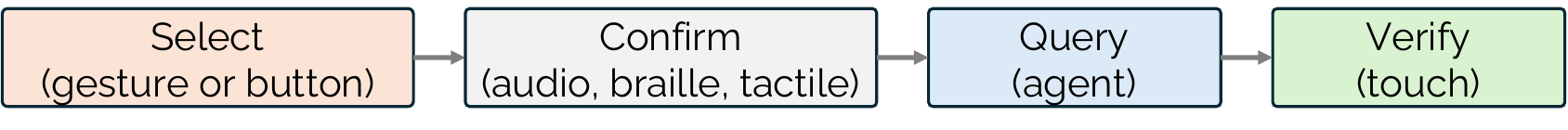}
      \vspace*{-4mm}
      \caption{The \textit{select, confirm, ask, verify} interaction pattern observed during free-form use. Users select data points, confirm selections through system feedback, query the agent, then verify responses by touch.}
      \label{fig:pattern}
      \vspace*{-4mm}
  \end{figure}

WS4 interaction patterns closely matched strategies co-designers had articulated during WS3, suggesting the co-designed approaches to reading and consuming charts were being internalized. Notably, co-designers consistently returned to touch after receiving the agent's response, tracing the referenced data to verify it --- extending the \textit{select, confirm, ask} pattern with a fourth step --- \textit{verify} (Fig.~\ref{fig:pattern}). 

\begin{table}[t]
  %%\centering
  \small
  \setlength{\tabcolsep}{2pt}
  \caption{WS4 interaction counts (left) and self-reported measures (right); Tracing: following chart elements, Comparison: relating positions across points or axes, Filtering: focusing strategy initiated via speech. NASA-TLX (1=Very Low to 7=Very High) -- lower is better except Performance; Naturalness (1=Strongly Disagree to 7=Strongly Agree) -- higher is better.\label{tab:ws4}
  \vspace*{-2mm}}
  \begin{minipage}[t]{0.52\columnwidth}
  \vspace{0pt}
  %%\centering
  %%\hspace*{-2mm}
  \begin{tabular}{lccc}
  \toprule & \textbf{D1} & \textbf{D2} & \textbf{D3}\\
  \midrule
  \multicolumn{4}{l}{\textit{Touch}}\\
  \hspace{1em}Tracing & 19 & 14 & 28\\
  \hspace{1em}Comparison & 5 & 5 & 10\\
  \hspace{1em}Braille reading & 4 & 16 & 1\\
  \midrule
  \multicolumn{4}{l}{\textit{Selection \& Navigation}}\\
  \hspace{1em}Button & 11 & 15 & 7\\
  \hspace{1em}Gesture & 10 & 8 & 6\\
  \midrule
  \multicolumn{4}{l}{\textit{Agent}}\\
  \hspace{1em}Questions & 3 & 2 & 3\\
  \hspace{1em}Filtering & 4 & 2 & 1\\
  \bottomrule
  \end{tabular}
  \end{minipage}
\hspace*{-2mm}
  \begin{minipage}[t]{0.46\columnwidth}
  \centering
  %%\small
  \setlength{\tabcolsep}{2pt}
  \vspace{0pt}
  \begin{tabular}{lccc}
  \toprule & \textbf{D1} & \textbf{D2} & \textbf{D3}\\
  \midrule
  \multicolumn{4}{l}{\textit{NASA-TLX}}\\
  \hspace{1em}Mental Demand    & 2 & 3 & 5\\
  \hspace{1em}Physical Demand  & 1 & 1 & 2\\
  \hspace{1em}Performance      & 7 & 7 & 6\\
  \hspace{1em}Effort           & 2 & 3 & 3\\
  \hspace{1em}Frustration      & 1 & 2 & 2\\
  \midrule
  \hspace{1em}Naturalness      & 7 & 6 & 6\\
  \bottomrule
  \end{tabular}
  \end{minipage}
  \vspace*{-4mm}
\end{table}

\subsubsection{Self-Reported Experience}
Following the free-form activity, co-designers reflected on their experience using the NASA-TLX dimensions and a question assessing the perceived naturalness of the system (\textit{``The way of interacting with the system felt natural''}, 7-point Likert scale) (Tab.~\ref{tab:ws4}). 

\textit{NASA-TLX:} Physical demand was deemed low by all the co-designers. Mental demand was deemed low by D1 and D2, but D1 noted that additional effort stemmed from hardware constraints, rather than interaction complexity: \textit{``The display resolution is the limit, not the system's capabilities.''} In contrast, D3 reported higher mental demand, noting that they were still learning the system's capabilities: \textit{``I still had to make deliberate interaction choices, to remember [for instance] `oh I can actually maybe ask this thing,' $\ldots$ but this will improve with use.''} Performance was deemed high by all the co-designers, with D2 noting that \textit{``using the system, I got to understand the trends and shape of the data well.''} Finally, D3 singled out latency as a source of friction.

\textit{Naturalness} was rated highly by all co-designers. D3 valued the breadth of multimodal interaction options: \textit{``I like the possibilities $\ldots$ use your hands and sometimes speech to confirm or verify; a very patient tutor.''} Even after the three-month gap between WS2 and WS3, they recalled how to ask the agent questions, select data points via gestures (though not the exact timing), and navigate the layered overview. By WS4, the co-designers used the full range of interactions unassisted. 

\subsubsection{Retrospective Reflections}\label{sec:reflection}
When reflecting on why they continued touch exploration after receiving an agent response, they described three main motivations: verification, sensemaking, and independence.

\textit{Verification:} D1 described doing \textit{``all three at once''}, but reflected that without the RTD, verification would not be possible: \textit{``Without the tactile display, I have to trust the agent, there's no certain way of verifying.''} D2 explained that they were \textit{``verifying the agent's accuracy and building [their own] understanding [through touch]''}, and noted, \textit{``I don't immediately trust anything with an LLM. I want to look at the actual data.''} Similarly, D3 spoke of prior experiences with AI: \textit{``I've experienced too many instances of AI not giving accurate information, so I don't trust it completely, and often verify it for accuracy.''} Even so, the co-designers emphasized that they would still want to explore by touch, even if the agent's responses were guaranteed to be correct (\eg\ computed from a spreadsheet, not an LLM). 

\textit{Sensemaking:} Touch was also described as central to building an overall understanding. D2 noted that \textit{``the richness of the information you get [through touch] is greater than a verbal response''}, and D1 added that \textit{``exploring the tactile representation of the data gives me a better overall understanding than a verbal summary, even if the exact details of particular data points are less clear.''} 

\textit{Autonomy and independence:} D3 described the agent as a last resort, preferring to resolve questions themselves through touch first: \textit{``Asking is the last resort, because I like to figure things out myself.''} D1 noted that \textit{``[touch] enables me to be independent, allowing me to consume data at my own pace.''} 

\section{Discussion \& Limitations}
\label{sec:discussion}
\subsection{The Complementary Nature of Touch and Conversation}\label{sec:complementary}
Touch and the agent took on distinct, complementary roles. Following Chundury \etal's recommendation~\cite{Chundury2026}, our work combined conversational speech interaction with tactile representations to address the limitations of each modality in isolation. Touch served as the primary sensemaking channel: co-designers traced to understand shape and trend, comparing and cross-referencing values against axes and other chart features. This aligns with our WOz study, where participants with tactile experience exhibited similar behavior. Our co-design process reveals why, showing that touch serves a broader role than independence.

In contrast, the conversational agent served as a data question-answering specialist, responding to queries that touch alone could not resolve. Its role evolved as familiarity grew: WS1 centered on basic value retrieval and extrema; WS2 saw use of trend analysis with deictic queries; WS3 shifted toward richer comparative analysis; and by WS4's more open-ended task (\S\ref{sec:freeform}), co-designers were attempting to push the agent beyond its current capabilities (\eg\ compound queries). In summary, touch supported the spatial understanding that the agent lacked, while the agent provided analytical depth that touch could not.

Our co-designers articulated reasons why they chose to anchor their exploration in touch rather than the agent: independence, verification, and sensemaking (\S\ref{sec:reflection}). Co-designers valued independence as a form of agency: the ability to build their own
understanding through touch, at their own pace, before (or instead of) asking the agent. This echoes Choe \etal's~\cite{Choe2025} finding that, when interpreting charts, sighted users with higher data literacy engaged more with charts and used LLMs as more of a specialist, whereas those with lower data literacy over-relied on the agent. Our BLV co-designers, who had both data and tactile experience, showed a similar pattern. Whether this holds for BLV users with less tactile and data experience remains an open question.

Touch also served as a method of verifying the agent's responses. Driven by a distrust of LLMs, co-designers reflected that without the plotted chart on the RTD, verification would not be possible (\S~\ref{sec:reflection}). Seo \etal~\cite{Seo2024b} illustrate this gap: although MAIDR supports tactile output in the form of Braille Unicode characters on single-line RTDs, without 2D spatial representations of charts, their BLV participants relied on other verification strategies, such as cross-referencing against other LLMs, or relying on background knowledge. A chart spatially rendered on an RTD offers something these strategies cannot: the ability to feel whether the agent's response matches the shape of the data.

At the same time, framing touch as only a verification method or means of maintaining independence undermines its broader role. Co-designers emphasized that they would still explore by touch even if the agent's responses were guaranteed to be correct, suggesting that the RTD provides something that speech fundamentally cannot: not just answers about the data, but a spatial understanding of it.

\subsection{Layered Presentation \& Responses}\label{sec:layered}
When a sighted user opens a chart, they can instantly survey its shape, peaks, and trends; a BLV user must move their hands and touch-read to build up an equivalent understanding, one feature at a time. On an RTD, this becomes even more difficult, due to its low resolution. This raises a fundamental design question: how to present charts to users on RTDs. A single continuous narration, akin to alt text or a transcriber's note~\cite{BANA2010guidelines}, proved mismatched: co-designers found it too long to follow while exploring by touch, with two channels competing for attention rather than reinforcing one another.

Their solution was a layered overview: chart components introduced one at a time, each rendered on the RTD with a spoken description. During workshops, co-designers insisted that the chart remain fully interactive within each layer, supporting touch exploration, gestures, and agent queries, and that layers be navigable, such that they could independently move back and forth through a chart. By WS4, the layered overview had evolved into what we now term a \textbf{layered presentation}: no longer just orientation and overview, but the primary framework in which co-designers explored and consumed their charts.

Each layer is displayed in isolation, creating a tight coupling between what is being heard and what is being felt: the spoken description narrates the component rendered on the RTD, and nothing else. Co-designers used this to focus their touch exploration, building spatial understanding without competing with tactile clutter. Zong \etal's~\cite{Zong2022} Olli breaks charts into individual components via a hierarchical tree navigable by screen readers; our layered presentation takes a similar approach in the tactile domain: components are introduced in a guided sequence that grounds the chart's structure in something that users can physically feel. D2's reversal, from \textit{``gimmicky''} to essential, illustrates how this coupling between touch and speech transformed a linear walkthrough into an interactive exploration structure.

The same challenge arises during agent interaction: longer responses overwhelm when users engage in touch exploration. Response segmentation addresses this by breaking agent responses into navigable one-sentence chunks, each synchronized with animated tactile highlights that draw attention to referenced data on the chart. This allows users to feel the data the agent is describing and pace the interaction themselves, verifying the agent's claims against the spatial layout of the chart and building their understanding as they go. Both the layered presentation and response segmentation serve the same purpose: structuring multimodal output so that touch and speech reinforce one another rather than compete for attention. 

Jiao \etal's~\cite{Jiao2025} tactile data comics share a step-by-step multimodal presentation on RTDs, but are strictly presentational; our layered presentation extends this into scaffolding for interactive data exploration, where co-designers could freely navigate layers or skip them entirely: D2 chose to skip the interpretive summary while still receiving the foundation they needed~\cite{Lundgard2022}.

The layered presentation was particularly valuable with unfamiliar chart types: all co-designers traversed back and forth through the scatterplot, building their conceptual understanding. D3 suggested the technique could generalize beyond charts to other tactile graphics, supported by Jiao's work~\cite{Jiao2025}, where step-by-step presentation improved comprehension across mathematics, maps, and literature.

\subsection{Beyond the Wizard: Comparing WOz to Co-Design}
In our earlier study~\cite{Reinders2024}, we used WOz to simulate an idealized system where a human wizard seamlessly coordinated touch and speech. Participants with more tactile experience preferred touch, with speech as a complementary channel. Our co-designers, each of whom had tactile experience, exhibited the same pattern, preferring to build understanding through touch before turning to the agent (\S~\ref{sec:complementary}). Our v1 prototype carried forward capabilities from the WOz study: gesture selection, free touch during speech, deictic queries, wake word activation, and a narrated chart overview (Tab.~\ref{tab:features}). Our co-designers retained some of these, replaced others, and introduced additional capabilities as new challenges emerged during the process of building a fully instantiated system that WOz had abstracted away:

\bipstart{When trust must be earned} In the WOz study, risk was invisible: as participants verbally articulated their intended interactions using think-aloud, the Wizard always correctly interpreted touch and speech, so the system never misunderstood. With a real conversational agent, infallibility is no longer guaranteed: speech transcriptions can misinterpret the query, and the LLM can hallucinate or otherwise produce incorrect results. With \name, co-designers developed their own verification strategies through touch, driven by a distrust of AI-generated responses (\S~\ref{sec:reflection}). They also rejected the free touch method identified in WOz, where participants would simultaneously touch points when asking their question, in favor of a sequential pattern: \textit{select, confirm, ask}, ensuring each step was grounded before querying the agent. This evolved into \textit{select, confirm, ask, verify} (\S~\ref{sec:freeform}), as co-designers used touch to verify the agent's responses against the chart.

\bipstart{When there is no guide} The Wizard served as a real-time guide, narrating chart overviews from prepared scripts. When instantiated, a single continuous narration (v1) proved difficult to follow, with co-designers unable to retain the information while simultaneously exploring the chart by touch. The layered presentation (\S~\ref{sec:layered}) emerged as the co-designed solution, evolving beyond the Wizard's orientation mechanism into the primary mode of exploring and consuming charts (\S~\ref{sec:freeform}).

\bipstart{When data must be traversed} WOz participants extracted values by gesturing specific points, or by asking the Wizard directly, but both methods required already knowing where a point was located. Stepping---moving through values one at a time using buttons (analogous to navigating with a screen reader)---was proposed as a faster, self-service alternative. It was particularly useful for precise interaction and recovery if gestures overshot, and for guided exploration where narrated values helped predict the shape of the data.

\bipstart{When encoding must be rethought} The WOz study used the Graphiti RTD, which supports raising pins to four heights. The current work used the Dot Pad RTD, which offers only binary output (raised or lowered). Without pin height as an encoding channel, co-designers shaped encodings within these constraints: designing distinct point symbols, connecting lines with clearance gaps, and fill textures for bar charts. However, this constraint also drove exploration of the encoding design space, including the intersection symbol and a feedback grammar that uses pin animation to distinguish user- and agent-initiated actions. This knowledge is specific to binary-pin RTDs, a constraint shared by the majority of the RTDs on the market (\eg\ the Monarch).

These challenges and solutions were also shaped by the longitudinal nature of the co-design process. Unlike our WOz study, which comprised single 90--120 minute sessions, this work spanned four workshops per co-designer over eight months, allowing co-designers to internalize interactions and refine preferences through repeated use---a depth our WOz study could not produce. 

\subsection{Design Recommendations}
Based on our findings, we contribute design guidance for researchers and developers building CTDIs:

\bipstart{Scaffold through layered presentation} Rather than a single continuous narration, introduce chart components sequentially~\cite{Jiao2025}, each rendered in isolation on the RTD with a spoken description, to reduce clutter. The same principle should extend to how the agent responds to questions: segment responses into navigable chunks, each synchronized with attention-drawing tactile feedback, so that users can feel the data being described at their own pace. Charts should always remain interactive, so that users can extend their sensemaking by selecting values and querying the agent.

\bipstart{Fuse touch and speech in agent interaction} Touch and speech complement one another: touch provides spatial grounding, the agent calculation and analytical depth~\cite{Chundury2026}. When users combine the two in deictic queries, the agent should acknowledge the touch context before answering so that users can confirm their input was correctly interpreted. Agent responses should be concise and answer-first, enabling verification grounded in the chart.

\bipstart{Ground each interaction step before the next} Each interaction should confirm the previous before moving on: select (via gesture or button), confirm the selection (via audio, Braille or tactile feedback), query the agent, then verify the response through touch. The pattern \textit{select, confirm, ask, verify} (Fig.~\ref{fig:pattern}) grounds each step in the data, ensuring queries have the correct context, and enabling users to check the agent's response against the chart.

\bipstart{Provide complementary navigation mechanisms} Different exploration goals require different navigation methods. Support gestures for selecting known points, and button stepping for traversing unknown points. Filtering can further aid exploration by reducing tactile clutter. 

\bipstart{Encode state using a tactile feedback grammar} Blinking pins can draw attention on RTDs~\cite{Holloway2022}, but when both user and agent can trigger highlights, users need to distinguish the source. Use distinct patterns for each: static highlights for gesture selections, animated highlights for agent references, and transitional animations for stepping that settle to static. Highlights should persist until dismissed by the user, affording them time to locate the highlighted component. The grammar should be consistent across chart types in order to reduce cognitive load.

\bipstart{Design tactile encodings for RTD constraints} Visual encoding conventions do not reliably transfer to tactile encoding~\cite{Marriott2026}, and conventions for printed tactile graphics do not necessarily transfer to RTDs due to their low resolution~\cite{Khalaila2026}. Encodings should be designed for touch discriminability with RTD pin arrays and be user-configurable, akin to how screen readers allow adjustment of speech rate and verbosity.

\subsection{Design Implications for Future RTD Devices}
CTDIs could be a game changer for accessible data visualization. At their core, though, they are constrained by RTD hardware itself. Our co-design process surfaced three hardware limitations of current pin-grid RTDs that manufacturers and researchers should consider:

\bpstart{Multi-height encoding} The Dot Pad and the Monarch RTD support only binary output, constraining encodings to a single pin height. The Graphiti RTD~\cite{Orbit} supports four heights, enabling data series and chart elements to be distinguished by elevation rather than symbols alone, but remains inaccessible at \$25,000 USD. As RTDs become more affordable, broader support for multiple heights would expand the design space of tactile encoding for data visualization on RTDs.
    
\bpstart{Multi-touch sensing} The Dot Pad lacks built-in touch sensing entirely; the Graphiti and Monarch support only single-finger touch detection, with the Monarch requiring users to lift other fingers when gesturing. We used an external finger tracker supporting both index fingers, but built-in multi-touch would enable interaction techniques such as pinch-to-zoom and multi-finger gestures for interactive data analysis.
    
\bpstart{Refresh under finger} The Dot Pad and Monarch use electromagnetic actuators~\cite{DotIncPatent} that cannot reliably raise pins against finger pressure, affecting Braille reading and tactile feedback alike. Reliable actuation under the finger would allow users to feel tactile feedback in place, making the search for highlighted elements easier.

Beyond these improvements, we see the need for tactile displays that move beyond binary, braille-pitch pin grids. Higher-resolution technologies or deformable surfaces would better support complex, denser charts. Further, we envision AI-native devices with the agent, speech processing, and rendering built into the RTD rather than distributed across a host computer and cloud services (as ours required), making multimodal interaction more portable and self-contained. 

\subsection{Limitations \& Future Work}\label{sec:limitations}
Several limitations of our work point to clear directions for future work. Our co-design involved three congenitally blind co-designers with varying levels of experience using tactile graphics and RTDs. This number is consistent with co-design practice in accessible visualization~\cite{Zong2022,Elavsky2023,Chundury2024,Seo2024}. Our findings may not generalize to BLV users with acquired blindness, less tactile experience, or residual vision, who may exhibit different usage patterns, and may rely more on the agent and less on independent touch verification. The layered presentation could give such users the confidence needed to engage with data through touch, while investigating interaction techniques better supporting agent-led exploration remains a clear opportunity.

We aimed to generate design knowledge, not to evaluate performance. While WS4 included free-form exploration with co-designers' own data, it was not a formal usability study. Future work should also include controlled and in-situ evaluation to assess how effectively users can consume and interpret data visualizations, and to measure the agent's accuracy and robustness, which we did not quantify.

Our co-design workshops utilized traditional RTDs, comprising binary pins on a fixed, braille-pitch grid. Hence some of our design knowledge, such as tactile encodings, may not transfer to future tactile-display paradigms. However, higher-level interaction principles, \eg\ the layered presentation and the \textit{select, confirm, ask, verify} pattern, are likely to be device-independent, and so carry over to future displays.

We explored a limited set of chart types (line, bar, stacked bar, scatterplot), all relatively low in data density; of these, scatterplots were the least resolved. More complex or denser charts will strain the low resolution of current RTDs, where adjacent or overlapping marks might be difficult to distinguish. We expect the layered presentation to be even more valuable in these cases, isolating components to keep each readable. Future work should expand our tactile encoding guidance to additional chart types. By WS4, co-designers were also pushing the agent's boundaries with compound queries requiring multiple operations, which Kim \etal~\cite{Kim2023} found prevalent among BLV users; support for compound queries grounded in touch would be a clear next step.

Co-designers also identified the need for interactive chart manipulation: charts that exceed the RTD's display require zooming and panning, but refreshing the entire display risks disorienting users who may lose spatial context~\cite{Reinders2024,zeng2014examples}. Maintaining spatial context during chart manipulation remains an open challenge for our future work, as does integration with existing tools like Power BI and Excel. 

%% if specified like this the section will be omitted in review mode
\acknowledgments{We gratefully acknowledge support from the Australian Research Council Discovery Projects scheme (DP220101221) and, in part, the Yonsei University Research Fund (2025-22-0099).}

\bibliographystyle{abbrv-doi-hyperref}

\bibliography{references}

\begin{thebibliography}{10}

\bibitem{alam2023seechart}
M.~Z.~I. Alam, S.~Islam, and E.~Hoque.
\newblock {SeeChart}: Enabling accessible visualizations through interactive natural language interface for people with visual impairments.
\newblock In {\em Proc.\ International Conference on Intelligent User Interfaces}, IUI '23, pp. 46––64. ACM, New York, 2023. \href{https://doi.org/10.1145/3581641.3584099}
{doi: {{%
10\hspace{.1pt}\discretionary{.}{%
}{.}\hspace{.4pt}1145\discretionary{/}{%
}{/}3581641\hspace{.1pt}\discretionary{.}{%
}{.}\hspace{.4pt}3584099}}}


\bibitem{APH}
{American Printing House for the Blind} - meet {Monarch}.
\newblock https://www.aph.org/meet-monarch/, 2023.

\bibitem{BANA2010guidelines}
{Braille Authority of North America}.
\newblock {\em Guidelines and Standards for Tactile Graphics}.
\newblock Braille Authority of North America, 2010.

\bibitem{brayda2019refreshable}
L.~Brayda, F.~Leo, C.~Baccelliere, C.~Vigini, and E.~Cocchi.
\newblock A refreshable tactile display effectively supports cognitive mapping followed by orientation and mobility tasks: A comparative multi-modal study involving blind and low-vision participants.
\newblock In {\em Proc.\ Multimedia for Accessible Human Computer Interfaces}, MAHCI '19, pp. 9--15. ACM, New York, 2019. \href{https://doi.org/10.1145/3347319.3356840}
{doi: {{%
10\hspace{.1pt}\discretionary{.}{%
}{.}\hspace{.4pt}1145\discretionary{/}{%
}{/}3347319\hspace{.1pt}\discretionary{.}{%
}{.}\hspace{.4pt}3356840}}}


\bibitem{Chen2025}
M.~K. Chen, I.~Pedraza~Pineros, A.~Satyanarayan, and J.~Zong.
\newblock Tactile {Vega-Lite}: Rapidly prototyping tactile charts with smart defaults.
\newblock In {\em Proc.\ CHI Conference on Human Factors in Computing Systems}, CHI '25,  art. no. 931,  23 pp. ACM, New York, 2025. \href{https://doi.org/10.1145/3706598.3714132}
{doi: {{%
10\hspace{.1pt}\discretionary{.}{%
}{.}\hspace{.4pt}1145\discretionary{/}{%
}{/}3706598\hspace{.1pt}\discretionary{.}{%
}{.}\hspace{.4pt}3714132}}}


\bibitem{Choe2025}
K.~Choe, C.~Lee, S.~Lee, J.~Song, A.~Cho, N.~W. Kim et al.
\newblock Enhancing data literacy on-demand: {LLMs} as guides for novices in chart interpretation.
\newblock {\em IEEE Transactions on Visualization and Computer Graphics}, 31(9):4712––4727, 2025. \href{https://doi.org/10.1109/TVCG.2024.3413195}
{doi: {{%
10\hspace{.1pt}\discretionary{.}{%
}{.}\hspace{.4pt}1109\discretionary{/}{%
}{/}TVCG\hspace{.1pt}\discretionary{.}{%
}{.}\hspace{.4pt}2024\hspace{.1pt}\discretionary{.}{%
}{.}\hspace{.4pt}3413195}}}


\bibitem{Chundury2026}
P.~Chundury, J.~B. Jordan, Y.~Reyazuddin, N.~Elmqvist, and J.~Lazar.
\newblock Sound, touch, or the full monty? {A} comparative study of accessible data exploration systems for blind users.
\newblock {\em ACM Transactions on Accessible Computing}, 19(1),  art. no. 1,  41 pp., 2026. \href{https://doi.org/10.1145/3798100}
{doi: {{%
10\hspace{.1pt}\discretionary{.}{%
}{.}\hspace{.4pt}1145\discretionary{/}{%
}{/}3798100}}}


\bibitem{Chundury2024}
P.~Chundury, Y.~Reyazuddin, J.~B. Jordan, J.~Lazar, and N.~Elmqvist.
\newblock {TactualPlot}: Spatializing data as sound using sensory substitution for touchscreen accessibility.
\newblock {\em IEEE Transactions on Visualization and Computer Graphics}, 30(1):836––846, 2024. \href{https://doi.org/10.1109/TVCG.2023.3326937}
{doi: {{%
10\hspace{.1pt}\discretionary{.}{%
}{.}\hspace{.4pt}1109\discretionary{/}{%
}{/}TVCG\hspace{.1pt}\discretionary{.}{%
}{.}\hspace{.4pt}2023\hspace{.1pt}\discretionary{.}{%
}{.}\hspace{.4pt}3326937}}}


\bibitem{DotInc}
{Dot Inc} - {Dot Pad} - the first tactile graphics display for the visually impaired.
\newblock https://pad.dotincorp.com/, 2022.

\bibitem{Elavsky2023}
F.~Elavsky, L.~Nadolskis, and D.~Moritz.
\newblock Data navigator: An accessibility-centered data navigation toolkit.
\newblock {\em IEEE Transactions on Visualization and Computer Graphics}, 30(1):803––813, 2024. \href{https://doi.org/10.1109/TVCG.2023.3327393}
{doi: {{%
10\hspace{.1pt}\discretionary{.}{%
}{.}\hspace{.4pt}1109\discretionary{/}{%
}{/}TVCG\hspace{.1pt}\discretionary{.}{%
}{.}\hspace{.4pt}2023\hspace{.1pt}\discretionary{.}{%
}{.}\hspace{.4pt}3327393}}}


\bibitem{farahani2023automatic}
A.~M. Farahani, P.~Adibi, M.~S. Ehsani, H.-P. Hutter, and A.~Darvishy.
\newblock Automatic chart understanding: A review.
\newblock {\em IEEE Access}, 11:76202––76221, 2023. \href{https://doi.org/10.1109/ACCESS.2023.3298050}
{doi: {{%
10\hspace{.1pt}\discretionary{.}{%
}{.}\hspace{.4pt}1109\discretionary{/}{%
}{/}ACCESS\hspace{.1pt}\discretionary{.}{%
}{.}\hspace{.4pt}2023\hspace{.1pt}\discretionary{.}{%
}{.}\hspace{.4pt}3298050}}}


\bibitem{GorniakEtAl2024Vizibility}
J.~Gorniak, Y.~Kim, D.~Wei, and N.~W. Kim.
\newblock {VizAbility}: Enhancing chart accessibility with {LLM}-based conversational interaction.
\newblock In {\em Proc.\ ACM Symposium on User Interface Software and Technology}, UIST '24,  art. no. 89,  19 pp. ACM, New York, 2024. \href{https://doi.org/10.1145/3654777.3676414}
{doi: {{%
10\hspace{.1pt}\discretionary{.}{%
}{.}\hspace{.4pt}1145\discretionary{/}{%
}{/}3654777\hspace{.1pt}\discretionary{.}{%
}{.}\hspace{.4pt}3676414}}}


\bibitem{Gyoshev2018}
S.~Gyoshev, D.~Karastoyanov, N.~Stoimenov, V.~Cantoni, L.~Lombardi, and A.~Setti.
\newblock Exploiting a graphical braille display for art masterpieces.
\newblock In {\em Proc.\ International Conference on Computers Helping People with Special Needs (ICCHP)}, pp. 237--245. Springer, 2018. \href{https://doi.org/10.1007/978-3-319-94274-2_33}
{doi: {{%
10\hspace{.1pt}\discretionary{.}{%
}{.}\hspace{.4pt}1007\discretionary{/}{%
}{/}978\discretionary{%
}{-}{-}3\discretionary{%
}{-}{-}319\discretionary{%
}{-}{-}94274\discretionary{%
}{-}{-}2\_33}}}


\bibitem{He2026}
T.~He, M.~McCracken, D.~Hajas, S.~Creem-Regehr, and A.~Lex.
\newblock Using tactile charts to support comprehension and learning of complex visualizations for blind and low-vision individuals.
\newblock {\em IEEE Transactions on Visualization and Computer Graphics}, 32(1):199––209, 2026. \href{https://doi.org/10.1109/TVCG.2025.3633874}
{doi: {{%
10\hspace{.1pt}\discretionary{.}{%
}{.}\hspace{.4pt}1109\discretionary{/}{%
}{/}TVCG\hspace{.1pt}\discretionary{.}{%
}{.}\hspace{.4pt}2025\hspace{.1pt}\discretionary{.}{%
}{.}\hspace{.4pt}3633874}}}


\bibitem{Holloway2022}
L.~Holloway, S.~Ananthanarayan, M.~Butler, M.~T. De~Silva, K.~Ellis, C.~Goncu et al.
\newblock Animations at your fingertips: Using a refreshable tactile display to convey motion graphics for people who are blind or have low vision.
\newblock In {\em Proc.\ ACM SIGACCESS Conference on Computers and Accessibility}, ASSETS '22,  art. no. 32,  16 pp. ACM, New York, 2022. \href{https://doi.org/10.1145/3517428.3544797}
{doi: {{%
10\hspace{.1pt}\discretionary{.}{%
}{.}\hspace{.4pt}1145\discretionary{/}{%
}{/}3517428\hspace{.1pt}\discretionary{.}{%
}{.}\hspace{.4pt}3544797}}}


\bibitem{holloway2024refreshable}
L.~Holloway, P.~Cracknell, K.~Stephens, M.~Fanshawe, S.~Reinders, K.~Marriott et al.
\newblock Refreshable tactile displays for accessible data visualisation.
\newblock {\em arXiv preprint arXiv:2401.15836}, 2024. \href{https://doi.org/10.48550/arXiv.2401.15836}
{doi: {{%
10\hspace{.1pt}\discretionary{.}{%
}{.}\hspace{.4pt}48550\discretionary{/}{%
}{/}arXiv\hspace{.1pt}\discretionary{.}{%
}{.}\hspace{.4pt}2401\hspace{.1pt}\discretionary{.}{%
}{.}\hspace{.4pt}15836}}}


\bibitem{holloway2022infosonics}
L.~M. Holloway, C.~Goncu, A.~Ilsar, M.~Butler, and K.~Marriott.
\newblock Infosonics: Accessible infographics for people who are blind using sonification and voice.
\newblock In {\em Proc.\ CHI Conference on Human Factors in Computing Systems}, CHI '22,  art. no. 480,  13 pp. ACM, New York, 2022. \href{https://doi.org/10.1145/3491102.3517465}
{doi: {{%
10\hspace{.1pt}\discretionary{.}{%
}{.}\hspace{.4pt}1145\discretionary{/}{%
}{/}3491102\hspace{.1pt}\discretionary{.}{%
}{.}\hspace{.4pt}3517465}}}


\bibitem{HoqueEtAl2022chart}
E.~Hoque, P.~Kavehzadeh, and A.~Masry.
\newblock Chart question answering: State of the art and future directions.
\newblock {\em Computer Graphics Forum}, 41(3):555--572, 2022. \href{https://doi.org/10.1111/cgf.14573}
{doi: {{%
10\hspace{.1pt}\discretionary{.}{%
}{.}\hspace{.4pt}1111\discretionary{/}{%
}{/}cgf\hspace{.1pt}\discretionary{.}{%
}{.}\hspace{.4pt}14573}}}


\bibitem{Jiao2025}
Y.~Jiao, R.~Sun, R.~Luo, X.~Yao, X.~She, K.~Hara et al.
\newblock Tactile data comics: Combining step-by-step presentation of tactile graphics with verbal narration for the blind and visually impaired.
\newblock In {\em Proc.\ ACM SIGACCESS Conference on Computers and Accessibility}, ASSETS '25,  art. no. 29,  14 pp. ACM, New York, 2025. \href{https://doi.org/10.1145/3663547.3746338}
{doi: {{%
10\hspace{.1pt}\discretionary{.}{%
}{.}\hspace{.4pt}1145\discretionary{/}{%
}{/}3663547\hspace{.1pt}\discretionary{.}{%
}{.}\hspace{.4pt}3746338}}}


\bibitem{jung2021communicating}
C.~Jung, S.~Mehta, A.~Kulkarni, Y.~Zhao, and Y.-S. Kim.
\newblock Communicating visualizations without visuals: Investigation of visualization alternative text for people with visual impairments.
\newblock {\em IEEE Transactions on Visualization and Computer Graphics}, 28(1):1095--1105, 2022. \href{https://doi.org/10.1109/TVCG.2021.3114846}
{doi: {{%
10\hspace{.1pt}\discretionary{.}{%
}{.}\hspace{.4pt}1109\discretionary{/}{%
}{/}TVCG\hspace{.1pt}\discretionary{.}{%
}{.}\hspace{.4pt}2021\hspace{.1pt}\discretionary{.}{%
}{.}\hspace{.4pt}3114846}}}


\bibitem{kavaz2023}
E.~Kavaz, A.~Puig, and I.~Rodríguez.
\newblock Chatbot-based natural language interfaces for data visualisation: A scoping review.
\newblock {\em Applied Sciences}, 13(12),  art. no. 7025, 2023. \href{https://doi.org/10.3390/app13127025}
{doi: {{%
10\hspace{.1pt}\discretionary{.}{%
}{.}\hspace{.4pt}3390\discretionary{/}{%
}{/}app13127025}}}


\bibitem{Khalaila2025}
A.~Khalaila and D.~Cashman.
\newblock Speculating a tactile grammar: Toward task-aligned chart design for non-visual perception.
\newblock In {\em Proc.\ ACM SIGACCESS Conference on Computers and Accessibility}, ASSETS '25,  art. no. 123,  4 pp. ACM, New York, 2025. \href{https://doi.org/10.1145/3663547.3759744}
{doi: {{%
10\hspace{.1pt}\discretionary{.}{%
}{.}\hspace{.4pt}1145\discretionary{/}{%
}{/}3663547\hspace{.1pt}\discretionary{.}{%
}{.}\hspace{.4pt}3759744}}}


\bibitem{Khalaila2026}
A.~Khalaila, L.~Harrison, N.~W. Kim, and D.~Cashman.
\newblock {``They Aren't Built for Me''}: An exploratory study of strategies for measurement of graphical primitives in tactile graphics.
\newblock {\em IEEE Transactions on Visualization and Computer Graphics}, 32(1):1--12, 2026. \href{https://doi.org/10.1109/TVCG.2025.3633881}
{doi: {{%
10\hspace{.1pt}\discretionary{.}{%
}{.}\hspace{.4pt}1109\discretionary{/}{%
}{/}TVCG\hspace{.1pt}\discretionary{.}{%
}{.}\hspace{.4pt}2025\hspace{.1pt}\discretionary{.}{%
}{.}\hspace{.4pt}3633881}}}


\bibitem{Kim2024}
H.~Kim, M.~Chung, E.~Kim, and Y.~Yoo.
\newblock Automatic video-to-audiotactile conversion of golf broadcasting on a refreshable pin array.
\newblock In {\em Proc.\ ACM Symposium on Spatial User Interaction}, SUI '24,  art. no. 15,  8 pp. ACM, New York, 2024. \href{https://doi.org/10.1145/3677386.3682092}
{doi: {{%
10\hspace{.1pt}\discretionary{.}{%
}{.}\hspace{.4pt}1145\discretionary{/}{%
}{/}3677386\hspace{.1pt}\discretionary{.}{%
}{.}\hspace{.4pt}3682092}}}


\bibitem{Kim2023}
J.~Kim, A.~Srinivasan, N.~W. Kim, and Y.-S. Kim.
\newblock Exploring chart question answering for blind and low vision users.
\newblock In {\em Proc.\ CHI Conference on Human Factors in Computing Systems}, CHI '23,  art. no. 828,  15 pp. ACM, New York, 2023. \href{https://doi.org/10.1145/3544548.3581532}
{doi: {{%
10\hspace{.1pt}\discretionary{.}{%
}{.}\hspace{.4pt}1145\discretionary{/}{%
}{/}3544548\hspace{.1pt}\discretionary{.}{%
}{.}\hspace{.4pt}3581532}}}


\bibitem{DotIncPatent}
J.~Y. Kim, J.~H. Kim, and H.~C. Park.
\newblock Information output device.
\newblock U.S. Patent 12,260,743. https://patents.google.com/patent/US12260743B2/, 2025.

\bibitem{kim2021accessible}
N.~W. Kim, S.~C. Joyner, A.~Riegelhuth, and Y.~Kim.
\newblock Accessible visualization: Design space, opportunities, and challenges.
\newblock {\em Computer Graphics Forum}, 40(3):173––188, 2021. \href{https://doi.org/10.1111/cgf.14298}
{doi: {{%
10\hspace{.1pt}\discretionary{.}{%
}{.}\hspace{.4pt}1111\discretionary{/}{%
}{/}cgf\hspace{.1pt}\discretionary{.}{%
}{.}\hspace{.4pt}14298}}}


\bibitem{Kim2019}
S.~Kim, Y.~Ryu, J.~Cho, and E.-S. Ryu.
\newblock Towards tangible vision for the visually impaired through {2D} multiarray braille display.
\newblock {\em Sensors}, 19(23),  art. no. 5319, 2019. \href{https://doi.org/10.3390/s19235319}
{doi: {{%
10\hspace{.1pt}\discretionary{.}{%
}{.}\hspace{.4pt}3390\discretionary{/}{%
}{/}s19235319}}}


\bibitem{Kim2025}
S.-g. Kim, A.~Kim, E.~Kim, M.~Chung, and Y.~Yoo.
\newblock Automatic accessible multimodal translation of graphics using a refreshable pin array.
\newblock In {\em Proc.\ ACM International Conference on Multimedia}, MM '25, pp. 7123––7132. ACM, New York, 2025. \href{https://doi.org/10.1145/3746027.3754586}
{doi: {{%
10\hspace{.1pt}\discretionary{.}{%
}{.}\hspace{.4pt}1145\discretionary{/}{%
}{/}3746027\hspace{.1pt}\discretionary{.}{%
}{.}\hspace{.4pt}3754586}}}


\bibitem{Kramer2010}
G.~Kramer, B.~Walker, T.~Bonebright, P.~Cook, J.~H. Flowers, N.~Miner et al.
\newblock Sonification report: Status of the field and research agenda.
\newblock Technical report, Department of Psychology, University of Nebraska, Lincoln, Nebraska, 2010.

\bibitem{lee2020reaching}
B.~Lee, E.~K. Choe, P.~Isenberg, K.~Marriott, and J.~Stasko.
\newblock Reaching broader audiences with data visualization.
\newblock {\em IEEE Computer Graphics and Applications}, 40(2):82--90, 2020. \href{https://doi.org/10.1109/MCG.2020.2968244}
{doi: {{%
10\hspace{.1pt}\discretionary{.}{%
}{.}\hspace{.4pt}1109\discretionary{/}{%
}{/}MCG\hspace{.1pt}\discretionary{.}{%
}{.}\hspace{.4pt}2020\hspace{.1pt}\discretionary{.}{%
}{.}\hspace{.4pt}2968244}}}


\bibitem{dagstuhl2023inclusive}
B.~Lee, K.~Marriott, D.~Szafir, and G.~Weber.
\newblock Inclusive data visualization ({Dagstuhl Seminar} 23252).
\newblock {\em Dagstuhl Reports}, 13(6):81--105, 2024. \href{https://doi.org/10.4230/DagRep.13.6.81}
{doi: {{%
10\hspace{.1pt}\discretionary{.}{%
}{.}\hspace{.4pt}4230\discretionary{/}{%
}{/}DagRep\hspace{.1pt}\discretionary{.}{%
}{.}\hspace{.4pt}13\hspace{.1pt}\discretionary{.}{%
}{.}\hspace{.4pt}6\hspace{.1pt}\discretionary{.}{%
}{.}\hspace{.4pt}81}}}


\bibitem{Lundgard2022}
A.~Lundgard and A.~Satyanarayan.
\newblock Accessible visualization via natural language descriptions: A four-level model of semantic content.
\newblock {\em IEEE Transactions on Visualization and Computer Graphics}, 28(1):1073––1083, 2022. \href{https://doi.org/10.1109/TVCG.2021.3114770}
{doi: {{%
10\hspace{.1pt}\discretionary{.}{%
}{.}\hspace{.4pt}1109\discretionary{/}{%
}{/}TVCG\hspace{.1pt}\discretionary{.}{%
}{.}\hspace{.4pt}2021\hspace{.1pt}\discretionary{.}{%
}{.}\hspace{.4pt}3114770}}}


\bibitem{Mankoff2010}
J.~Mankoff, G.~R. Hayes, and D.~Kasnitz.
\newblock Disability studies as a source of critical inquiry for the field of assistive technology.
\newblock In {\em Proc.\ ACM SIGACCESS Conference on Computers and Accessibility}, ASSETS '10, pp. 3--10. ACM, New York, 2010. \href{https://doi.org/10.1145/1878803.1878807}
{doi: {{%
10\hspace{.1pt}\discretionary{.}{%
}{.}\hspace{.4pt}1145\discretionary{/}{%
}{/}1878803\hspace{.1pt}\discretionary{.}{%
}{.}\hspace{.4pt}1878807}}}


\bibitem{ManzoniEtAl2024MapIO}
M.~Manzoni, S.~Mascetti, D.~Ahmetovic, R.~Crabb, and J.~M. Coughlan.
\newblock {MapIO}: A gestural and conversational interface for tactile maps.
\newblock {\em IEEE Access}, 13:84038--84056, 2025. \href{https://doi.org/10.1109/ACCESS.2025.3566286}
{doi: {{%
10\hspace{.1pt}\discretionary{.}{%
}{.}\hspace{.4pt}1109\discretionary{/}{%
}{/}ACCESS\hspace{.1pt}\discretionary{.}{%
}{.}\hspace{.4pt}2025\hspace{.1pt}\discretionary{.}{%
}{.}\hspace{.4pt}3566286}}}


\bibitem{Marriott2026}
K.~Marriott, M.~Butler, L.~Holloway, W.~Jolley, B.~Lee, B.~Maguire et al.
\newblock {From Vision to Touch}: Bridging visual and tactile principles for accessible data representation.
\newblock {\em IEEE Transactions on Visualization and Computer Graphics}, 32(1):659––669, 2026. \href{https://doi.org/10.1109/TVCG.2025.3634254}
{doi: {{%
10\hspace{.1pt}\discretionary{.}{%
}{.}\hspace{.4pt}1109\discretionary{/}{%
}{/}TVCG\hspace{.1pt}\discretionary{.}{%
}{.}\hspace{.4pt}2025\hspace{.1pt}\discretionary{.}{%
}{.}\hspace{.4pt}3634254}}}


\bibitem{marriott2021inclusive}
K.~Marriott, B.~Lee, M.~Butler, E.~Cutrell, K.~Ellis, C.~Goncu et al.
\newblock Inclusive data visualization for people with disabilities: A call to action.
\newblock {\em Interactions}, 28(3):47––51, 2021. \href{https://doi.org/10.1145/3457875}
{doi: {{%
10\hspace{.1pt}\discretionary{.}{%
}{.}\hspace{.4pt}1145\discretionary{/}{%
}{/}3457875}}}


\bibitem{Motoyoshi2018}
T.~Motoyoshi, S.~Mizushima, K.~Sawai, T.~Tamamoto, H.~Masuta, K.~Koyanagi et al.
\newblock Prototype development of a shape presentation system using linear actuators.
\newblock In {\em Proc.\ International Conference on Computers Helping People with Special Needs (ICCHP)}, pp. 226--230. Springer, 2018. \href{https://doi.org/10.1007/978-3-319-94274-2_31}
{doi: {{%
10\hspace{.1pt}\discretionary{.}{%
}{.}\hspace{.4pt}1007\discretionary{/}{%
}{/}978\discretionary{%
}{-}{-}3\discretionary{%
}{-}{-}319\discretionary{%
}{-}{-}94274\discretionary{%
}{-}{-}2\_31}}}


\bibitem{Namdev2015}
R.~K. Namdev and P.~Maes.
\newblock An interactive and intuitive {STEM} accessibility system for the blind and visually impaired.
\newblock In {\em Proc.\ International Conference on Pervasive Technologies Related to Assistive Environments (PETRA)}, pp. 1––7. ACM, New York, 2015. \href{https://doi.org/10.1145/2769493.2769502}
{doi: {{%
10\hspace{.1pt}\discretionary{.}{%
}{.}\hspace{.4pt}1145\discretionary{/}{%
}{/}2769493\hspace{.1pt}\discretionary{.}{%
}{.}\hspace{.4pt}2769502}}}


\bibitem{Ohshima2021football}
H.~Ohshima, M.~Kobayashi, and S.~Shimada.
\newblock Development of blind football play-by-play system for visually impaired spectators: Tangible sports.
\newblock In {\em Ext.\ Abstr.\ of the CHI Conference on Human Factors in Computing Systems}, CHI EA '21,  art. no. 209,  6 pp. ACM, New York, 2021. \href{https://doi.org/10.1145/3411763.3451737}
{doi: {{%
10\hspace{.1pt}\discretionary{.}{%
}{.}\hspace{.4pt}1145\discretionary{/}{%
}{/}3411763\hspace{.1pt}\discretionary{.}{%
}{.}\hspace{.4pt}3451737}}}


\bibitem{Orbit}
{Orbit Research} - {Graphiti} - a breakthrough in non-visual access to all forms of graphical information.
\newblock http://www.orbitresearch.com/product/graphiti/, 2016.

\bibitem{Park2016}
T.~Park, J.~Jung, and J.~Cho.
\newblock A method for automatically translating print books into electronic {Braille} books.
\newblock {\em Science China Information Sciences}, 59(7),  art. no. 072101, 2016. \href{https://doi.org/10.1007/s11432-016-5575-z}
{doi: {{%
10\hspace{.1pt}\discretionary{.}{%
}{.}\hspace{.4pt}1007\discretionary{/}{%
}{/}s11432\discretionary{%
}{-}{-}016\discretionary{%
}{-}{-}5575\discretionary{%
}{-}{-}z}}}


\bibitem{Ramoa2025}
G.~Ram\^{o}a and K.~M\"{u}ller.
\newblock Leveraging dynamic audio-tactile uis to assist visually impaired users in exploring line charts through tactile graphic readers.
\newblock In {\em Proc.\ ACM SIGACCESS Conference on Computers and Accessibility}, ASSETS '25,  art. no. 28,  15 pp. ACM, New York, 2025. \href{https://doi.org/10.1145/3663547.3746398}
{doi: {{%
10\hspace{.1pt}\discretionary{.}{%
}{.}\hspace{.4pt}1145\discretionary{/}{%
}{/}3663547\hspace{.1pt}\discretionary{.}{%
}{.}\hspace{.4pt}3746398}}}


\bibitem{Reinders2023}
S.~Reinders, S.~Ananthanarayan, M.~Butler, and K.~Marriott.
\newblock Designing conversational multimodal {3D} printed models with people who are blind.
\newblock In {\em Proc.\ ACM Designing Interactive Systems Conference}, DIS '23, pp. 2172––2188. ACM, New York, 2023. \href{https://doi.org/10.1145/3563657.3595989}
{doi: {{%
10\hspace{.1pt}\discretionary{.}{%
}{.}\hspace{.4pt}1145\discretionary{/}{%
}{/}3563657\hspace{.1pt}\discretionary{.}{%
}{.}\hspace{.4pt}3595989}}}


\bibitem{Reinders2020}
S.~Reinders, M.~Butler, and K.~Marriott.
\newblock {``Hey Model!''} -- natural user interactions and agency in accessible interactive {3D} models.
\newblock In {\em Proc.\ CHI Conference on Human Factors in Computing Systems}, CHI '20, pp. 1--13. ACM, New York, 2020. \href{https://doi.org/10.1145/3313831.3376145}
{doi: {{%
10\hspace{.1pt}\discretionary{.}{%
}{.}\hspace{.4pt}1145\discretionary{/}{%
}{/}3313831\hspace{.1pt}\discretionary{.}{%
}{.}\hspace{.4pt}3376145}}}


\bibitem{Reinders2025}
S.~Reinders, M.~Butler, and K.~Marriott.
\newblock {``It Brought the Model to Life''}: Exploring the embodiment of multimodal {I3Ms} for people who are blind or have low vision.
\newblock In {\em Proc.\ CHI Conference on Human Factors in Computing Systems}, CHI '25,  art. no. 367,  19 pp. ACM, New York, 2025. \href{https://doi.org/10.1145/3706598.3713158}
{doi: {{%
10\hspace{.1pt}\discretionary{.}{%
}{.}\hspace{.4pt}1145\discretionary{/}{%
}{/}3706598\hspace{.1pt}\discretionary{.}{%
}{.}\hspace{.4pt}3713158}}}


\bibitem{Reinders2024}
S.~Reinders, M.~Butler, I.~Zukerman, B.~Lee, L.~Qu, and K.~Marriott.
\newblock {When Refreshable Tactile Displays Meet Conversational Agents}: Investigating accessible data presentation and analysis with touch and speech.
\newblock {\em IEEE Transactions on Visualization and Computer Graphics}, 31(1):864––874, 2025. \href{https://doi.org/10.1109/TVCG.2024.3456358}
{doi: {{%
10\hspace{.1pt}\discretionary{.}{%
}{.}\hspace{.4pt}1109\discretionary{/}{%
}{/}TVCG\hspace{.1pt}\discretionary{.}{%
}{.}\hspace{.4pt}2024\hspace{.1pt}\discretionary{.}{%
}{.}\hspace{.4pt}3456358}}}


\bibitem{Reinders2026}
S.~Reinders, M.~Zaib, M.~Butler, B.~Lee, I.~Zukerman, L.~Qu et al.
\newblock Supporting multimodal data interaction on refreshable tactile displays: An architecture to combine touch and conversational {AI}.
\newblock In {\em Proc.\ IEEE Pacific Visualization Conference (PacificVis)}, pp. 73--78. IEEE, 2026. \href{https://doi.org/10.1109/PacificVis68791.2026.00013}
{doi: {{%
10\hspace{.1pt}\discretionary{.}{%
}{.}\hspace{.4pt}1109\discretionary{/}{%
}{/}PacificVis68791\hspace{.1pt}\discretionary{.}{%
}{.}\hspace{.4pt}2026\hspace{.1pt}\discretionary{.}{%
}{.}\hspace{.4pt}00013}}}


\bibitem{rowell2003world}
J.~Rowell and S.~Ungar.
\newblock The world of touch: An international survey of tactile maps. {P}art 1: Production.
\newblock {\em British Journal of Visual Impairment}, 21(3):98--104, 2003. \href{https://doi.org/10.1177/026461960302100303}
{doi: {{%
10\hspace{.1pt}\discretionary{.}{%
}{.}\hspace{.4pt}1177\discretionary{/}{%
}{/}026461960302100303}}}


\bibitem{Seo2024b}
J.~Seo, S.~S. Kamath, A.~Zeidieh, S.~Venkatesh, and S.~McCurry.
\newblock {MAIDR} meets {AI}: Exploring multimodal {LLM}-based data visualization interpretation by and with blind and low-vision users.
\newblock In {\em Proc.\ ACM SIGACCESS Conference on Computers and Accessibility}, ASSETS '24,  art. no. 57,  31 pp. ACM, New York, 2024. \href{https://doi.org/10.1145/3663548.3675660}
{doi: {{%
10\hspace{.1pt}\discretionary{.}{%
}{.}\hspace{.4pt}1145\discretionary{/}{%
}{/}3663548\hspace{.1pt}\discretionary{.}{%
}{.}\hspace{.4pt}3675660}}}


\bibitem{Seo2024}
J.~Seo, Y.~Xia, B.~Lee, S.~McCurry, and Y.~J. Yam.
\newblock {MAIDR}: Making statistical visualizations accessible with multimodal data representation.
\newblock In {\em Proc.\ CHI Conference on Human Factors in Computing Systems}, CHI '24,  art. no. 211,  22 pp. ACM, New York, 2024. \href{https://doi.org/10.1145/3613904.3642730}
{doi: {{%
10\hspace{.1pt}\discretionary{.}{%
}{.}\hspace{.4pt}1145\discretionary{/}{%
}{/}3613904\hspace{.1pt}\discretionary{.}{%
}{.}\hspace{.4pt}3642730}}}


\bibitem{SharifEtAlCHI2022}
A.~Sharif, O.~H. Wang, A.~T. Muongchan, K.~Reinecke, and J.~O. Wobbrock.
\newblock {VoxLens}: Making online data visualizations accessible with an interactive {JavaScript} plug-in.
\newblock In {\em Proc.\ CHI Conference on Human Factors in Computing Systems}, CHI '22,  art. no. 478,  19 pp. ACM, New York, 2022. \href{https://doi.org/10.1145/3491102.3517431}
{doi: {{%
10\hspace{.1pt}\discretionary{.}{%
}{.}\hspace{.4pt}1145\discretionary{/}{%
}{/}3491102\hspace{.1pt}\discretionary{.}{%
}{.}\hspace{.4pt}3517431}}}


\bibitem{siu2022supporting}
A.~Siu, G.~S-H~Kim, S.~O'Modhrain, and S.~Follmer.
\newblock Supporting accessible data visualization through audio data narratives.
\newblock In {\em Proc.\ CHI Conference on Human Factors in Computing Systems}, CHI '22,  art. no. 476,  19 pp. ACM, New York, 2022. \href{https://doi.org/10.1145/3491102.3517678}
{doi: {{%
10\hspace{.1pt}\discretionary{.}{%
}{.}\hspace{.4pt}1145\discretionary{/}{%
}{/}3491102\hspace{.1pt}\discretionary{.}{%
}{.}\hspace{.4pt}3517678}}}


\bibitem{Smiley2025}
J.~Smiley, S.~Ananthanarayan, L.~M. Holloway, M.~Butler, and T.~Dwyer.
\newblock {MagnePins}: A modular, affordable, and {DIY} refreshable braille and tactile display.
\newblock In {\em Proc.\ ACM Symposium on User Interface Software and Technology}, UIST '25,  art. no. 69,  12 pp. ACM, New York, 2025. \href{https://doi.org/10.1145/3746059.3747692}
{doi: {{%
10\hspace{.1pt}\discretionary{.}{%
}{.}\hspace{.4pt}1145\discretionary{/}{%
}{/}3746059\hspace{.1pt}\discretionary{.}{%
}{.}\hspace{.4pt}3747692}}}


\bibitem{Thompson2023}
J.~R. Thompson, J.~J. Martinez, A.~Sarikaya, E.~Cutrell, and B.~Lee.
\newblock {Chart Reader}: Accessible visualization experiences designed with screen reader users.
\newblock In {\em Proc.\ CHI Conference on Human Factors in Computing Systems}, CHI '23,  art. no. 802,  18 pp. ACM, New York, 2023. \href{https://doi.org/10.1145/3544548.3581186}
{doi: {{%
10\hspace{.1pt}\discretionary{.}{%
}{.}\hspace{.4pt}1145\discretionary{/}{%
}{/}3544548\hspace{.1pt}\discretionary{.}{%
}{.}\hspace{.4pt}3581186}}}


\bibitem{Wobbrock2009}
J.~O. Wobbrock, M.~R. Morris, and A.~D. Wilson.
\newblock User-defined gestures for surface computing.
\newblock In {\em Proc.\ CHI Conference on Human Factors in Computing Systems}, CHI '09, pp. 1083--1092. ACM, New York, 2009. \href{https://doi.org/10.1145/1518701.1518866}
{doi: {{%
10\hspace{.1pt}\discretionary{.}{%
}{.}\hspace{.4pt}1145\discretionary{/}{%
}{/}1518701\hspace{.1pt}\discretionary{.}{%
}{.}\hspace{.4pt}1518866}}}


\bibitem{Yang2021}
W.~Yang, J.~Huang, R.~Wang, W.~Zhang, H.~Liu, and J.~Xiao.
\newblock A survey on tactile displays for visually impaired people.
\newblock {\em IEEE Transactions on Haptics}, 14(4):712––721, 2021. \href{https://doi.org/10.1109/TOH.2021.3085915}
{doi: {{%
10\hspace{.1pt}\discretionary{.}{%
}{.}\hspace{.4pt}1109\discretionary{/}{%
}{/}TOH\hspace{.1pt}\discretionary{.}{%
}{.}\hspace{.4pt}2021\hspace{.1pt}\discretionary{.}{%
}{.}\hspace{.4pt}3085915}}}


\bibitem{Zeng2015}
L.~Zeng, M.~Miao, and G.~Weber.
\newblock Interactive audio-haptic map explorer on a tactile display.
\newblock {\em Interacting with Computers}, 27(4):413--429, 2015. \href{https://doi.org/10.1093/iwc/iwu006}
{doi: {{%
10\hspace{.1pt}\discretionary{.}{%
}{.}\hspace{.4pt}1093\discretionary{/}{%
}{/}iwc\discretionary{/}{%
}{/}iwu006}}}


\bibitem{zeng2014examples}
L.~Zeng, G.~Weber, I.~Zoller, P.~Lotz, T.~A. Kern, J.~Reisinger et al.
\newblock Examples of haptic system development.
\newblock In C.~Hatzfeld and T.~A. Kern, eds., {\em Engineering Haptic Devices: A Beginner's Guide}, pp. 525--554. Springer, 2nd ed., 2014. \href{https://doi.org/10.1007/978-1-4471-6518-7_14}
{doi: {{%
10\hspace{.1pt}\discretionary{.}{%
}{.}\hspace{.4pt}1007\discretionary{/}{%
}{/}978\discretionary{%
}{-}{-}1\discretionary{%
}{-}{-}4471\discretionary{%
}{-}{-}6518\discretionary{%
}{-}{-}7\_14}}}


\bibitem{Zhao2025}
Y.~Zhao, Y.~Zhang, Y.~Zhang, X.~Zhao, J.~Wang, Z.~Shao et al.
\newblock {LEVA}: Using large language models to enhance visual analytics.
\newblock {\em IEEE Transactions on Visualization and Computer Graphics}, 31(3):1830––1847, 2025. \href{https://doi.org/10.1109/TVCG.2024.3368060}
{doi: {{%
10\hspace{.1pt}\discretionary{.}{%
}{.}\hspace{.4pt}1109\discretionary{/}{%
}{/}TVCG\hspace{.1pt}\discretionary{.}{%
}{.}\hspace{.4pt}2024\hspace{.1pt}\discretionary{.}{%
}{.}\hspace{.4pt}3368060}}}


\bibitem{Zong2022}
J.~Zong, C.~Lee, A.~Lundgard, J.~Jang, D.~Hajas, and A.~Satyanarayan.
\newblock Rich screen reader experiences for accessible data visualization.
\newblock {\em Computer Graphics Forum}, 41(3):15––27, 2022. \href{https://doi.org/10.1111/cgf.14519}
{doi: {{%
10\hspace{.1pt}\discretionary{.}{%
}{.}\hspace{.4pt}1111\discretionary{/}{%
}{/}cgf\hspace{.1pt}\discretionary{.}{%
}{.}\hspace{.4pt}14519}}}


\bibitem{Zong2024}
J.~Zong, I.~Pedraza~Pineros, M.~K. Chen, D.~Hajas, and A.~Satyanarayan.
\newblock Umwelt: Accessible structured editing of multi-modal data representations.
\newblock In {\em Proc.\ CHI Conference on Human Factors in Computing Systems}, CHI '24,  art. no. 46,  20 pp. ACM, New York, 2024. \href{https://doi.org/10.1145/3613904.3641996}
{doi: {{%
10\hspace{.1pt}\discretionary{.}{%
}{.}\hspace{.4pt}1145\discretionary{/}{%
}{/}3613904\hspace{.1pt}\discretionary{.}{%
}{.}\hspace{.4pt}3641996}}}


\end{thebibliography}

%\appendix % You can use the `hideappendix` class option to skip everything after \appendix
%\crefalias{section}{appendix} % this is to make sure that cleverref switches to referring to Appx. X from here on

\end{document}